\documentclass{article} 
\usepackage{nips14submit_e,times}
\usepackage{hyperref}
\usepackage{url}

\usepackage{wrapfig}
\usepackage{multirow}
\usepackage{graphicx}
\usepackage{tikz}
\usepackage{amsthm}
\usepackage{amsfonts}
\usepackage{amsmath}

\usepackage[noend]{algorithmic}
\usepackage{algorithm}
\usepackage{caption}
\usepackage{commands}
\usepackage{maddison_commands}
\usepackage{3dplot} 
\usetikzlibrary{arrows,calc,trees,shapes,decorations.pathreplacing}

\title{\OurAlgorithm}

\author{
Chris J. Maddison\\
Dept. of Computer Science\\
University of Toronto\\
\texttt{cmaddis@cs.toronto.edu} \\
\And
Daniel Tarlow, Tom Minka \\
Microsoft Research \\
\texttt{\{dtarlow,minka\}@microsoft.com}
}

\nipsfinalcopy 

\begin{document}

\maketitle
\vspace{-10pt}
\begin{abstract}
\vspace{-3pt}
The problem of drawing samples from a discrete distribution can be
converted into a discrete optimization problem \cite{Papandreou11,
tarlow2012randomized,hazan2012partition,ermon2013embed}.  In this
work, we show how sampling from a continuous
distribution can be converted into an optimization problem over 
continuous space.
Central to the method 
is a stochastic process recently described in mathematical statistics
that we call the
\emph{\ourprocess}. We present a new construction of the \ourprocess~and \emph{\OurAlgorithm},
a practical generic sampling algorithm that searches for the maximum of a Gumbel process
using \astar~search. 
We analyze the correctness and convergence time of \OurAlgorithm~and demonstrate empirically that it
makes more efficient use of bound and likelihood evaluations than the most closely
related  adaptive rejection sampling-based algorithms.
\end{abstract}

\vspace{-7pt}
\section{Introduction}
\vspace{-7pt}

Drawing samples from arbitrary probability distributions is a core
problem in statistics and machine learning. Sampling methods are used
widely when training, evaluating, and predicting with probabilistic
models.  In this work, we introduce a generic sampling algorithm
that returns exact independent samples from a distribution of interest. This line of work is important as we seek to include probabilistic models as
subcomponents in larger systems, and as we seek to build probabilistic
modelling tools that are usable by non-experts; in these cases, guaranteeing the quality of inference is 
highly desirable.
There are a range of existing approaches for exact sampling.
Some are specialized to specific distributions \cite{papandreou2010gaussian}, but exact generic methods are 
based either on (adaptive) rejection sampling \cite{gilks1992adaptive,dymetman2012algorithm, mansinghka2009exact}
or  Markov Chain Monte Carlo (MCMC) methods where
convergence to the stationary distribution can be guaranteed
\cite{propp1996exact,mira2001perfect,mitha2003perfect}. 

This work approaches the problem from a different perspective. Specifically, it
is inspired by an algorithm for sampling from a discrete distribution
that is known as the \gumbeltrick. 
The algorithm works by adding independent
Gumbel perturbations to each configuration of a discrete
negative energy function and returning the argmax configuration of the 
perturbed negative energy function. The result is an exact sample from
the corresponding Gibbs distribution.
Previous work \cite{Papandreou11,hazan2012partition} has used this property to motivate
samplers based on optimizing random energy functions but has been forced
to resort to approximate sampling due to the fact that in structured output
spaces, exact sampling appears to require instantiating exponentially many
Gumbel perturbations.

Our first key observation is that we can apply the
\gumbeltrick~without instantiating all of
the (possibly exponentially many) Gumbel perturbations.
The same basic idea then allows us to extend the
\gumbeltrick~to continuous spaces where there will be infinitely many
independent perturbations.  Intuitively, for any given random energy function,
there are many perturbation values that are irrelevant to determining the argmax
so long as we have an upper bound on their values.
We will show how to instantiate the relevant ones and bound the irrelevant ones,
allowing us to find the argmax --- and thus an exact sample.

There are a number of challenges that must be overcome along the way,
which are addressed in this work. First, what does it mean to
independently perturb space in a way analogous to perturbations in the
\gumbeltrick? We introduce the \ourprocess, a 
special case of a stochastic process recently defined in mathematical
statistics \cite{malmberg2013random}, which generalizes the notion of perturbation over space.  Second, we need a method for
working with a
\ourprocess~that does not require instantiating infinitely many
random variables. This leads to our novel construction of the Gumbel
process, which draws perturbations according to a top-down ordering
of their values. Just as the stick breaking construction of
the Dirichlet process gives insight into algorithms for the Dirichlet
process, our construction gives insight into algorithms for the Gumbel
process. We demonstrate this by developing \ouralgorithm, 
which leverages the construction to draw samples from
arbitrary continuous distributions. We study the
relationship between \ouralgorithm~and adaptive rejection sampling-based
methods and identify a key 
difference that leads to more efficient use of bound and likelihood
computations. We investigate
the behaviour of \ouralgorithm~on a variety of illustrative and challenging
problems.

\vspace{-7pt}
\section{The \OurProcess}
\label{sec:intuition}
\vspace{-7pt}

The \gumbeltrick~is an algorithm for sampling
from a categorical distribution over classes $i
\in \{1 \dotdotdot n\}$ with probability proportional to
$\exp(\negenergy(i))$.  The algorithm proceeds by adding independent
Gumbel-distributed noise to the log-unnormalized mass $\negenergy(i)$ and returns the optimal
class of the perturbed distribution. In more detail, $\G \sim \GumbelDist(m)$ 
is a Gumbel with location $m$ if $\prob(\G \leq \g) = \exp(-\exp(-\g  + m))$.
The \gumbeltrick~follows from the structure of Gumbel distributions and
basic properties of order statistics;
if $\G(i)$ are i.i.d.\ $\GumbelDist(0)$, then $\argmax_{i} \left\{\G(i) +
\negenergy(i) \right\} \sim \exp(\negenergy(i))/\sum_i
\exp(\negenergy(i))$.
Further,
for any $B
\subseteq \{1 \dotdotdot n\}$
\begin{align}
\max_{i \in B} \left\{\G(i) + \negenergy(i) \right\} &\sim \GumbelDist\left(\log \sum_{i \in B} \exp(\negenergy(i))\right)\label{eq:maxstability}\\
\argmax_{i \in B} \left\{\G(i) + \negenergy(i) \right\} &\sim \frac{\exp(\negenergy(i))}{\sum_{i \in B} \exp(\negenergy(i))}\label{eq:luceschoice}
\end{align}
\eqref{eq:maxstability} is known as \emph{max-stability}---the highest
order statistic of a sample of independent Gumbels also has a Gumbel
distribution with a location that is the log partition
function  \cite{gumbel}. \eqref{eq:luceschoice} is a consequence of the fact that Gumbels satisfy Luce's choice
axiom \cite{YellottJr1977109}. Moreover, the max and argmax are independent random variables, see Appendix for proofs.

We would like to generalize the interpretation to continuous distributions as
maximizing over the perturbation of a density $p(\x) \propto \exp(\negenergy(\x))$ on $\realsd$.
The perturbed density should have properties analogous to the discrete case, namely
that the max in $B \subseteq \realsd$ should be distributed as
$\GumbelDist(\log \int_{\x \in B} \exp(\negenergy(\x)))$ and the distribution of the argmax in $B$ should be distributed $\propto \indicate{\x \in B}\exp(\negenergy(\x))$. The Gumbel process is a generalization satisfying these properties.

\begin{deff}
\label{def:gup}
Adapted from \cite{malmberg2013random}. Let $\volume(B)$ be a sigma-finite measure on sample space $\Omega$, $B \subseteq \Omega$ measurable, and $\G_{\volume}(B)$ a random variable. $\GUP_{\volume} = \{\G_{\volume}(B) \given B \subseteq \Omega\}$ is a \ourprocess, if
\begin{enumerate}
\item  \label{def:marginals} ({marginal distributions}) $\G_{\volume}(B) \sim \GumbelDist\left(\log \volume(B)\right).$
\item \label{def:independence} ({independence of disjoint sets}) $\G_{\volume}(B) \;\perp\; \G_{\volume}(B^c).$
\item \label{def:consistency} ({consistency constraints}) for measurable $A , B \subseteq \Omega$, then
\begin{align*}
\G_{\volume}(A \cup B) = \max(\G_{\volume}(A), \G_{\volume}(B)).
\end{align*} 
\end{enumerate}
\end{deff}

The marginal distributions condition ensures that the \ourprocess~ satisfies the requirement on the max. The consistency requirement ensures that a realization of a Gumbel process is consistent across space. Together with the independence these ensure the argmax requirement: the argmax of a Gumbel process restricted to $B^{\prime}$ is distributed according to the probability distribution that is proportional to the sigma-finite measure $\mu$ restricted to $B^{\prime}$. In particular, let $\bar{\mu}( B \given B^{\prime}) = \mu(B \cap B^{\prime})/\mu(B^{\prime})$ be the probability distribution proportional to $\mu$.
If $\G_{\volume}(B)$ is the optimal value of some
perturbed density restricted to $B$, then the event that the optima over $\Omega$ is
contained in $B$ is equivalent to the event that $\G_{\volume}(B) \geq
\G_{\volume}(B^c)$.
The conditions ensure
 that $\prob(\G_{\volume}(B) \geq \G_{\volume}(B^c)) = \bar{\mu}(B \given \Omega)$
\cite{malmberg2013random}. 
Thus, for example we can
use the \ourprocess~ for a continuous measure $\mu(B) = \int_{\x \in B} \exp(\negenergy(\x))$ on $\realsd$ to model a perturbed density
function where the optimum is distributed $\propto \exp(\phi(x))$. Notice that this definition is a generalization of the finite case; if $\Omega$ is finite, then the collection
$\GUP_{\volume}$ corresponds exactly to maxes over subsets of independent Gumbels.

\vspace{-7pt}
\section{Top-Down Construction for the \OurProcess}
\vspace{-7pt}

While \cite{malmberg2013random} defines and constructs a general class of stochastic
processes that include the \ourprocess, the construction that proves
their existence gives little insight into how to execute a
continuous version of the \gumbeltrick.  Here we give an alternative algorithmic
construction that will form the foundation of our practical sampling algorithm.
In this section we assume $\log \measure(\Omega)$ can be computed
tractably; this assumption will be lifted in \secref{sec:algs}. To
explain the construction, we consider the discrete case as an
introductory example.

\begin{wrapfigure}[19]{r}[0pt]{0pt}
\begin{minipage}[t]{.55\textwidth}
\vspace{-22pt}
  \begin{algorithm}[H]
  \footnotesize{
   \caption{\footnotesize{Top-Down Construction}}
   \label{alg:gup}
   \begin{algorithmic}
   \INPUT{sample space $\Omega$, sigma-finite measure $\mu(B)$}
   \STATE $(B_1, Q) \gets (\Omega, \Queue)$
   \STATE $\G_{1} \sim \GumbelDist(\log \measure(\Omega)) $
   \STATE $\X_1  \sim \bar{\mu}( \cdot \given \Omega)$
   \STATE $Q.push(1)$
   \STATE $k \gets 1$
   \WHILE {!$Q.empty()$}
   	\STATE $p \gets Q.pop()$
	\STATE $L, R \gets partition(B_p - \{\X_{p}\})$
	\FOR {$C \in \{L, R\}$}
		\IF {$C \neq \emptyset$}
			\STATE $k \gets k + 1$
			\STATE $B_k \gets C$
		   	\STATE $\G_{k} \sim \TruncGumbelDist(\log \measure(B_k), \G_p) $
		  	\STATE $\X_{k} \sim \bar{\mu}( \cdot \given B_k)$
			\STATE $Q.push(k)$
			\STATE {\bf yield} $\, (G_k, X_k)$
		\ENDIF
	\ENDFOR
   \ENDWHILE
  \end{algorithmic}
  }
\end{algorithm}
\end{minipage}
\end{wrapfigure}

Suppose $\G_{\volume}(i) \sim \GumbelDist(\negenergy(i))$ is
a set of independent Gumbel random variables for $i \in \{1 \dotdotdot n\}$.
It would be straightforward 
to sample the variables then build a heap of the $\G_{\volume}(i)$ values and 
also have heap nodes store the index $i$ associated with their value.
Let $B_i$ be the set of indices that appear in the
subtree rooted at the node with index $i$. A property of the heap is that the root $(\G_{\volume}(i), i)$ pair
is the max and argmax of the set of Gumbels with index in $B_i$.
The key idea of our construction is to sample the independent set of
random variables by instantiating this heap from root to leaves.
That is, we will first sample the root node, which is the global 
max and argmax, then we will recurse, sampling the root's two children
conditional upon the root. At the end, we will have sampled a heap full of
values and indices; reading off the value associated with each index
will yield a draw of independent Gumbels from the target distribution.

We sketch an inductive argument. For the base
case, sample the max and its index $i^*$ using their distributions that
we know from
\eqref{eq:maxstability} and \eqref{eq:luceschoice}.
Note the max and argmax are independent. Also let $B_{i^*}=\{0, \ldots, n-1\}$ be
the set of all indices.
Now, inductively, suppose have sampled a partial heap and would like to 
recurse downward starting at $(\G_{\volume}(p), p)$.
Partition the remaining indices to be sampled $B_p - \{p\}$
into two subsets $L$ and $R$ and let $l \in
L$ be the left argmax and $r \in R$ be the right argmax.
Let $[\ge \!\!p]$ be the indices that have been sampled already.
 Then
\vspace{-2pt}
\begin{align}
&p\left(\G_{\volume}(l)=\g_l, \G_{\volume}(r)=\g_r, \{ \G_{\volume}(k)=\g_k\}_{k \in [\ge p]} \given [\ge \!\!p]\right)\\
 \propto
& p\left(\max_{i \in L} \G_{\volume}(i) \! = \! \g_l \right)
  p\left(\max_{i \in R} \G_{\volume}(i) \! =\! \g_r \right) 
\prod_{k \in [\ge p]} p_k(\G_{\volume}(k) = g_k) \indicate{g_k \ge g_{\mathcal{L}(k)} \wedge g_k \ge g_{\mathcal{R}(k)}}
\nonumber
\end{align}
where $\mathcal{L}(k)$ and $\mathcal{R}(k)$ denote the left and right children of $k$
and the constraints should only be applied amongst nodes $[\ge \!\! p] \cup \{l, r\}$.
This implies
\begin{align}
&p\left(\G_{\volume}(l)=\g_l, \G_{\volume}(r)=\g_r \given  \{ \G_{\volume}(k)=\g_k\}_{k \in [\ge p]}, [\ge \!\!p] \right) \nonumber\\
&\propto p\left(\max_{i \in L} \G_{\volume}(i) =\g_l \right)   
p\left(\max_{i \in R} \G_{\volume}(i) =\g_r \right) \indicate{\g_{p} > \g_{l}}\indicate{\g_{p} > \g_{r}} \label{eq:truncjust}.
\end{align}
\eqref{eq:truncjust} is the joint density of two 
independent Gumbels truncated at $\G_{\volume}(p)$. We could
sample the children maxes and argmaxes by sampling the independent Gumbels
in $L$ and $R$ respectively and computing their maxes,
rejecting those that exceed the known value of $\G_{\volume}(p)$. 
Better, the truncated Gumbel distributions can be
sampled efficiently via CDF inversion\footnote{$\G \sim
\TruncGumbelDist(\negenergy, b)$ if $G$ has CDF $\exp(-\exp(-\min(\g,
b) + \negenergy))/\exp(-\exp(-b + \negenergy))$. To sample
efficiently, return $\G = -\log(\exp(-b - \gamma +
\negenergy) -\log(U)) - \gamma + \negenergy$ where $U \sim
\Uniform[0,1]$.}, and the independent argmaxes within $L$ and $R$
can be sampled using \eqref{eq:luceschoice}. Note that any
choice of partitioning strategy for 
$L$ and $R$ leads to the
same distribution over the set of Gumbel values.

The basic structure of this top-down sampling procedure allows us to deal with infinite spaces; we can still generate an infinite descending heap of Gumbels and locations as if we had made a heap from an infinite list.
The algorithm (which appears as \algref{alg:gup}) 
begins by sampling the optimal value $\G_1 \sim \GumbelDist(\log \volume(\Omega))$ over sample space $\Omega$ and its location $\X_1 \sim \bar{\mu}(\cdot \given \Omega)$. $\X_1$ is removed from
the sample space and the remaining sample space is partitioned into $L$ and $R$. 
The optimal Gumbel values for $L$ and $R$
are sampled from a Gumbel with location log measure of their respective sets, but truncated at $\G_1$.
The locations are sampled independently
from their sets, and the procedure recurses.  As in the discrete case, this yields a stream of $(\G_k, \X_k)$ pairs, which we can think of as being nodes in a heap of the $\G_k$'s.

If $\G_{\volume}(\x)$
is the value of the perturbed negative energy at $\x$, 
then \algref{alg:gup} instantiates this function at countably many points
by setting $\G_{\volume}(\X_k) = \G_k$. 
In the discrete case we eventually sample the complete perturbed
density, but in the continuous case we simply generate an infinite
stream of locations and values.
The sense in which \algref{alg:gup} constructs
a \ourprocess~is that the collection
$\{\max \{\G_k \given \X_k \in B\} \given B \subseteq \Omega\}$
satisfies Definition \ref{def:gup}. 
The intuition should be provided by the introductory argument;
a full proof appears in the Appendix.
An important note is that because $\G_k$'s
are sampled in descending order along a path in the tree, when the first $\X_k$ lands
in set $B$, the value of $\max \{\G_k \given \X_k \in B\}$ will not
change as the algorithm continues.

\begin{wrapfigure}[36]{R}[0pt]{0pt}
\noindent\begin{minipage}[t]{.5\textwidth}
\vspace{-50pt}

\begin{figure}[H]
\centering
\includegraphics[width=.85\columnwidth]{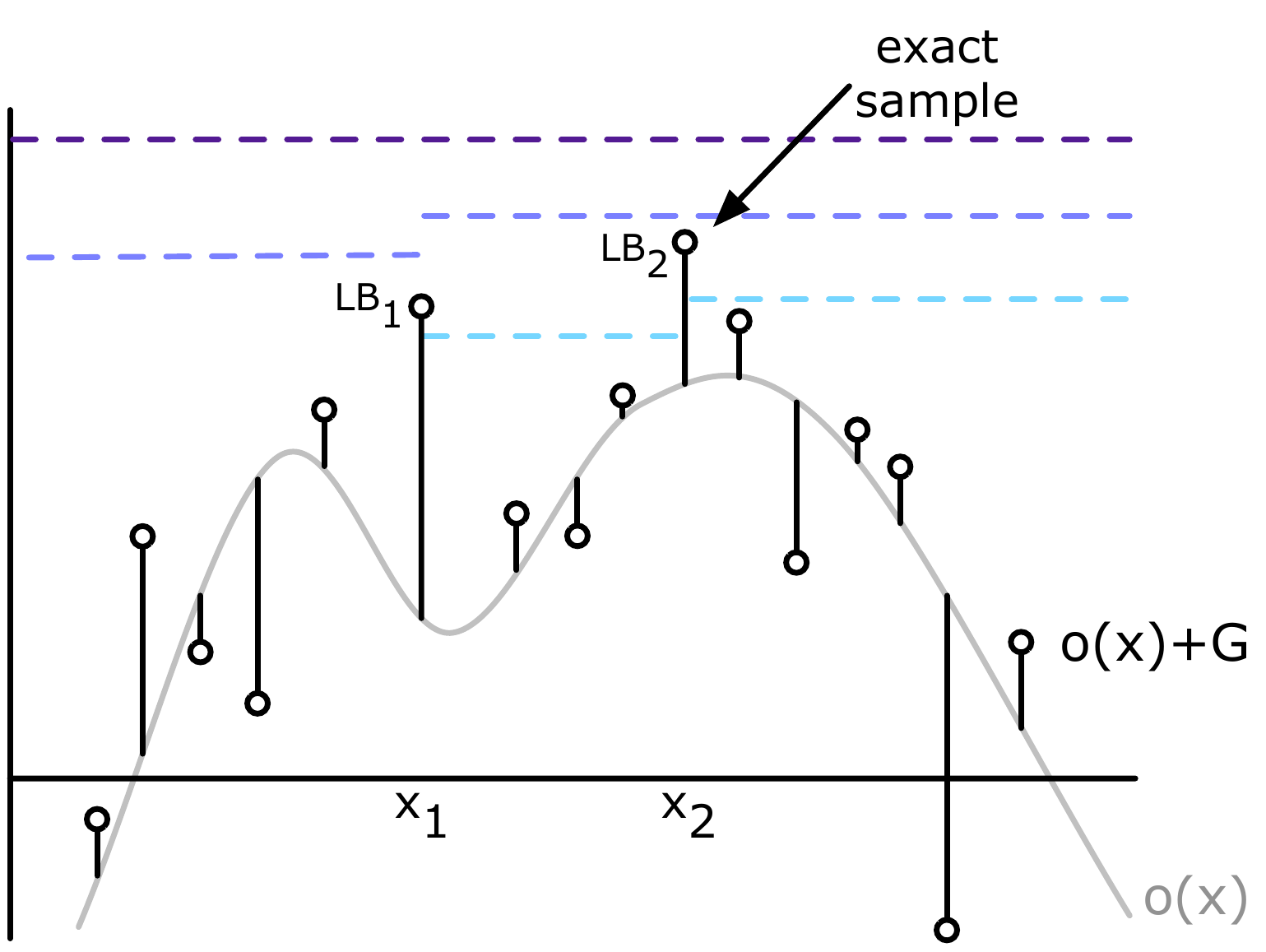}
\vspace{-8pt}
\caption{
\footnotesize{Illustration of \ouralgorithm.}
}
\label{fig:astar-illustration}
\end{figure}
\vspace{-4pt}

\vspace{-17pt}
\begin{algorithm}[H]
\footnotesize{
\caption{\footnotesize{\OurAlgorithm}}
    \label{alg:astar}
    \begin{algorithmic}
\INPUT log density $i(x)$, difference $o(x)$, bounding function $M(B)$, and $partition$
\STATE $(LB, \; \X^* , \; k)\gets ( -\infty,\;  \text{null} ,\; 1)$
\STATE $Q \gets \PriorityQueue$
\STATE $\G_1 \sim \GumbelDist(\log \measurei(\realsd))$
\STATE $\X_1 \sim \exp(i(\x)) / \measurei(\realsd))$
\STATE $\bound_1 \gets M(\realsd)$
\STATE $Q.pushWithPriority(1, \G_1 + \bound_1)$
\WHILE {$!Q.empty()$ and $LB < Q.topPriority()$}
	\STATE $p \gets Q.popHighest()$
		\STATE $LB_p \gets \G_p + o(\X_p)$
		\IF{$LB < LB_p$}
			\STATE $LB \gets LB_p$
			\STATE $X^* \gets \X_p$
		\ENDIF
	\STATE $L, R \gets partition(B_p, \X_p)$ 
	\FOR{$C \in \{ L, R\}$}
		\IF{$C \neq \emptyset$}
		\STATE $k \gets k + 1$
		\STATE $B_k \gets C$
		\STATE $\G_{k}  \sim \TruncGumbelDist(\log \measurei(B_k), \G_p)$
		\STATE $\X_{k}  \sim \indicate{\x \in B_k} \exp(\i(\x)) / \measurei(B_k)$ 	
		\IF{$LB < \G_{k} + \bound_p$}
			\STATE $\bound_{k} \gets M(B_k)$
			\IF {$LB < \G_{k} + \bound_{k}$}
				\STATE $Q.pushWithPriority(k, \G_k + \bound_k)$
			\ENDIF
		\ENDIF
		\ENDIF
	\ENDFOR
\ENDWHILE
\OUTPUT $(LB, \X^*)$
    \end{algorithmic}
}
\end{algorithm}
\end{minipage}
\end{wrapfigure}

\vspace{-7pt}
\section{\OurAlgorithm} 
\label{sec:algs}
\vspace{-7pt}

The Top-Down construction is not executable in
general, because it assumes $\log \measure (\Omega)$ can be computed
efficiently.
\ouralgorithm~is an
algorithm that executes the \gumbeltrick~without this assumption by exploiting properties of the Gumbel process. Henceforth \ouralgorithm~refers exclusively to the continuous version.

\ouralgorithm~is possible because we can transform
one Gumbel process into another by adding the difference in their log
densities.  Suppose we have two continuous measures $\measure(B) =
\int_{x \in B} \exp(\negenergy(x))$ and $\measurei(B) = \int_{x \in B}
\exp(i(x))$.  Let pairs $(\G_k, \X_k)$ be draws from the Top-Down
construction for $\GUP_{\measurei}$.  If $\o(\x) = \negenergy(\x)
-\i(\x)$ is bounded, then we can recover $\GUP_{\measure}$ by adding
the difference $o(\X_k)$ to every $\G_k$; i.e., $\{\max \{\G_k +
\o(\X_k) \given \X_k \in B\} \given B \subseteq \realsd\}$ is a
\ourprocess~with measure $\volume$. As an example, if $\measurei$ were
a prior and $o(x)$ a bounded log-likelihood, then we could simulate
the Gumbel process corresponding to the posterior by adding $o(\X_k)$
to every $\G_k$ from a run of the construction for $\measurei$.

This ``linearity'' allows us to decompose a target log density
function into a tractable $i(x)$ and boundable
$o(x)$. The tractable component is analogous to the proposal
distribution in a rejection sampler. \ouralgorithm~searches
for $\argmax\{\G_k + \o(\X_k)\}$ within the heap of $(\G_k, \X_k)$ pairs from the Top-Down construction of $\GUP_{\measurei}$. The search 
is an \astar~procedure: nodes in the search tree correspond to 
increasingly refined regions
in space, and the search is guided by upper and lower bounds that are
computed for each region.
Lower bounds for region $B$ come from drawing the max $\G_k$
and argmax $\X_k$ of $\GUP_{\measurei}$ within $B$ and
evaluating $\G_k + o(\X_k)$.  Upper bounds come from the fact that
\begin{align*}
& \max \{\G_k + \o(\X_k) \given \X_k \in B\}
 \le
\max \{\G_k \given \X_k \in B\} + M(B), \nonumber
\end{align*}
where $M(B)$ is a bounding function for a region, $M(B) \geq o(x)$ for all $\x \in B$. $M(B)$ is not random and can be implemented using methods from e.g., convex duality or interval analysis. The first term on the RHS is
the $\G_k$ value used in the lower bound.

The algorithm appears in \algref{alg:astar} and an execution
is illustrated in \figref{fig:astar-illustration}. The algorithm
begins with a global upper bound (dark blue dashed).  $\G_1$ and $\X_1$ are sampled, and the first lower bound
$LB_1=\G_1 + o(\X_1)$ is computed. Space is split, upper
bounds are computed for the new children regions (medium blue dashed), 
and the new nodes are put on the queue. 
The region with highest upper bound
is chosen, the maximum Gumbel in the region, $(\G_2, \X_2)$, is sampled, and $LB_2$ is computed.
The current region is split at $\X_2$ (producing light blue dashed bounds), 
after which $LB_2$ is greater than the upper bound
for any region on the queue, so $LB_2$ is guaranteed to be the max over the infinite tree of $\G_k + \o(\X_k)$. Because $\max \{\G_k + \o(\X_k) \given \X_k \in B\}$ is a \ourprocess~with measure $\measure$, this means that $\X_2$ is an exact sample from 
$p(x) \propto \exp(\negenergy(x)))$ 
and $LB_2$ is an exact sample from $\GumbelDist(\log \measure(\realsd))$. Proofs of termination and correctness are in the Appendix.

\textbf{\OurAlgorithm~Variants. }
There are several variants of \ouralgorithm. 
When more than one sample
is desired, bound information can be reused across runs of the
sampler. In particular, suppose we have a partition of $\realsd$ with
bounds on $o(\x)$ for each region. \ouralgorithm~could use this by
running a search independently for each region and returning the max
Gumbel. The maximization can be done lazily by using \astar~search,
only expanding nodes in regions that are needed to determine
the global maximum.  The second variant trades bound
computations for likelhood computations by drawing more
than one sample from the auxiliary Gumbel process at each node in the
search tree. In this way, more lower bounds are computed (costing more
likelihood evaluations), but if this leads to better lower bounds,
then more regions of space can be pruned, leading to fewer bound
evaluations.  Finally, an interesting special case of
\ouralgorithm~can be implemented when $o(\x)$ is unimodal in 1D. In
this case, at every split of a parent node, one child can immediately
be pruned, so the ``search'' can be executed without a queue.  It
simply maintains the currently active node and drills down until it
has provably found the optimum.

\vspace{-7pt}
\section{Comparison to Rejection Samplers}
\label{sec:comparison}
\vspace{-7pt}

Our first result relating \ouralgorithm~to rejection sampling is
that if the same global bound $M = M(\realsd)$ is used at all nodes within
\ouralgorithm, then the runtime of \ouralgorithm~is
equivalent to that of standard rejection sampling. That is, the
distribution over the number of iterations is distributed as a
Geometric distribution with rate parameter
$\measure(\realsd)/(\exp(M)\measurei(\realsd))$. A proof 
is in the Appendix as part of the proof of
termination. 

When bounds are refined, \ouralgorithm~bears similarity to adaptive rejection
sampling-based algorithms. In particular, while it appears only to have been
applied in discrete domains, \osstar~\cite{dymetman2012algorithm} is a
general class of adaptive rejection sampling methods that maintain
piecewise bounds on the target distribution. If piecewise constant
bounds are used (henceforth we assume \osstar~uses only constant bounds) the procedure
can be described as follows:\ at
each step, (1) a region $B$ with bound $M(B)$ is sampled with probability proportional
to $\measurei(B)\exp(M(B))$, (2) a point is drawn from the proposal distribution restricted to the chosen region;
(3) standard accept/rejection computations are performed using the regional bound,
and (4) if the point is rejected, a region is chosen to be split into
two, and new bounds are computed for the two regions that were created
by the split. This process repeats until a point is accepted.

Steps (2) and (4) are
performed identically in \astar~when sampling argmax Gumbel
locations and when splitting a parent node.
A key difference is how regions are chosen in step (1). In \osstar, a
region is drawn according to
volume of the region under the proposal. Note that piece selection could be implemented using the \gumbeltrick,
in which case we would choose the piece with maximum $\G_B + M(B)$ where $\G_B \sim \GumbelDist(\log \measurei(B))$.
In \ouralgorithm~the region with highest upper bound is chosen, where
the upper bound is $\G_B + M(B)$. The difference is that
$\G_B$ values are reset after each rejection in \osstar, while they 
persist in \ouralgorithm~until a sample is returned. 

The effect of the difference is that \ouralgorithm~more tightly
couples together where the accepted sample will be and which regions
are refined. Unlike \osstar, it can go so far as to prune a region from the search, meaning
there is zero probability that the returned sample will be from that region,
and that region will never be refined further.
\osstar, on the other hand, is blind towards where the sample that will eventually
be accepted comes from and will on average waste more computation refining regions
that ultimately are not useful in drawing the sample. In experiments, we will see that
\astar~consistently dominates \osstar, refining the function less while also
using fewer likelihood evaluations. This is possible because the persistence
inside \ouralgorithm~focuses the refinement on the regions that are important
for accepting the current sample.

\vspace{-7pt}
\section{Experiments}
\label{sec:experiments}
\vspace{-7pt}

There are three main aims in this section. First, understand
the empirical behavior of \ouralgorithm~as parameters of the inference problem
and $o(x)$ bounds vary.
Second, demonstrate generality by showing that
\ouralgorithm~algorithms can be instantiated in just a few lines of
model-specific code by expressing $o(x)$ symbolically, and then using
a branch and bound library to automatically compute bounds.
Finally, compare to \osstar~and an MCMC method (slice sampling).
In all experiments, regions in the
search trees are hyper rectangles (possibly with infinite extent); to split a region $A$, 
choose the dimension with the largest side length and split the dimension at
the sampled $\X_k$ point.

\begin{figure}
\begin{tabular}{ccc}
\includegraphics[width=.207\columnwidth]{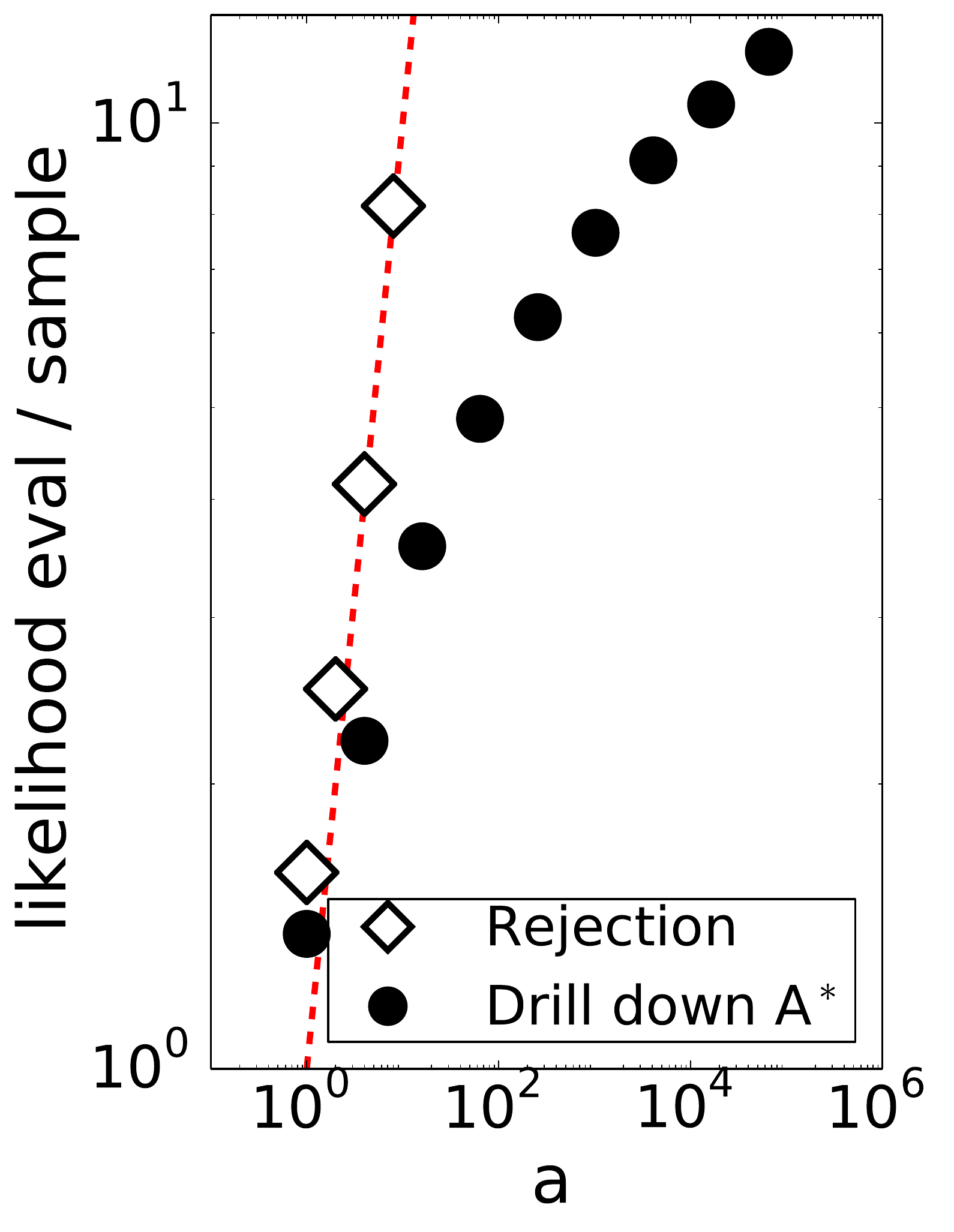} &
\includegraphics[width=.263\columnwidth]{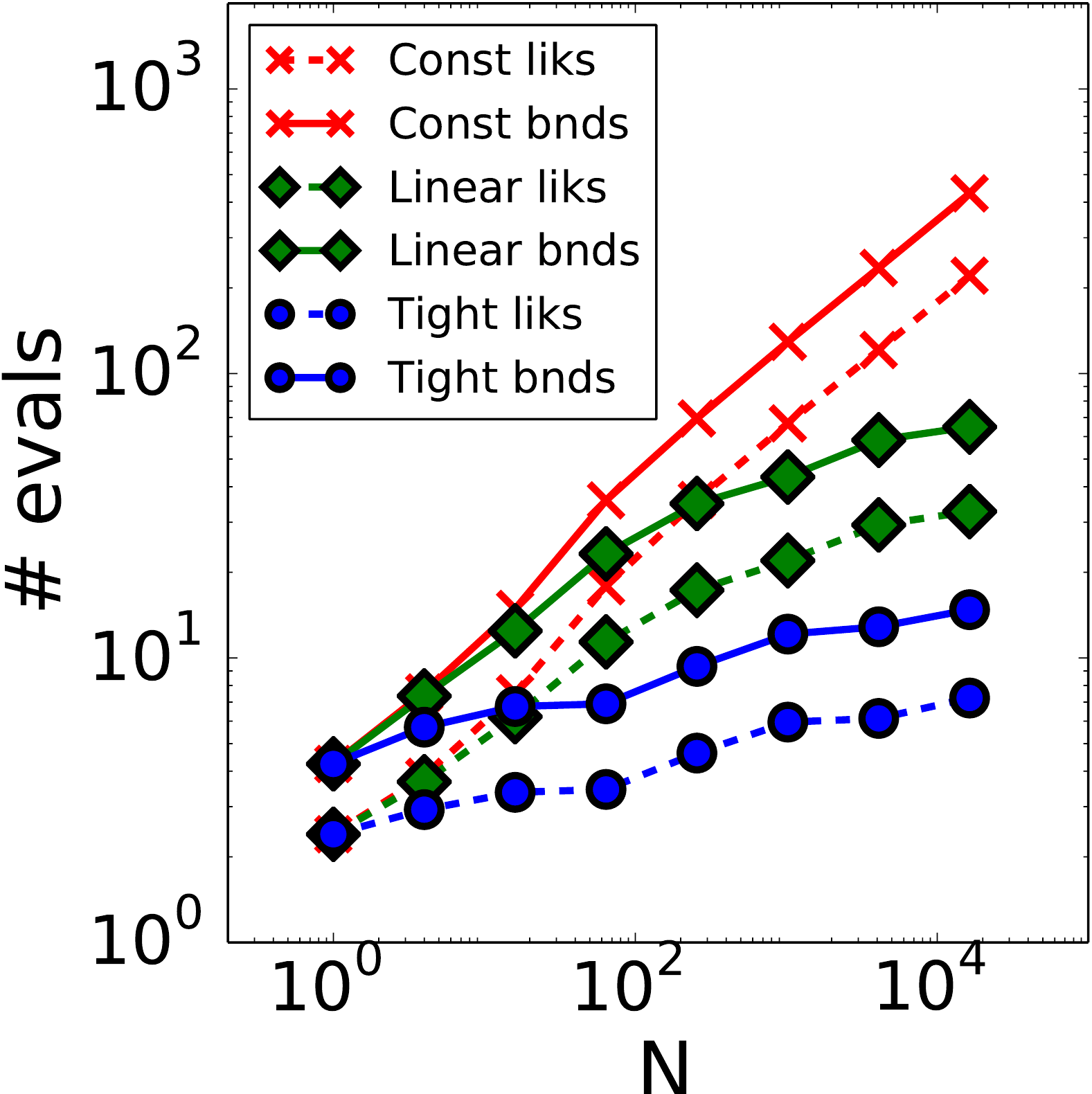} & 
\includegraphics[width=.445\columnwidth]{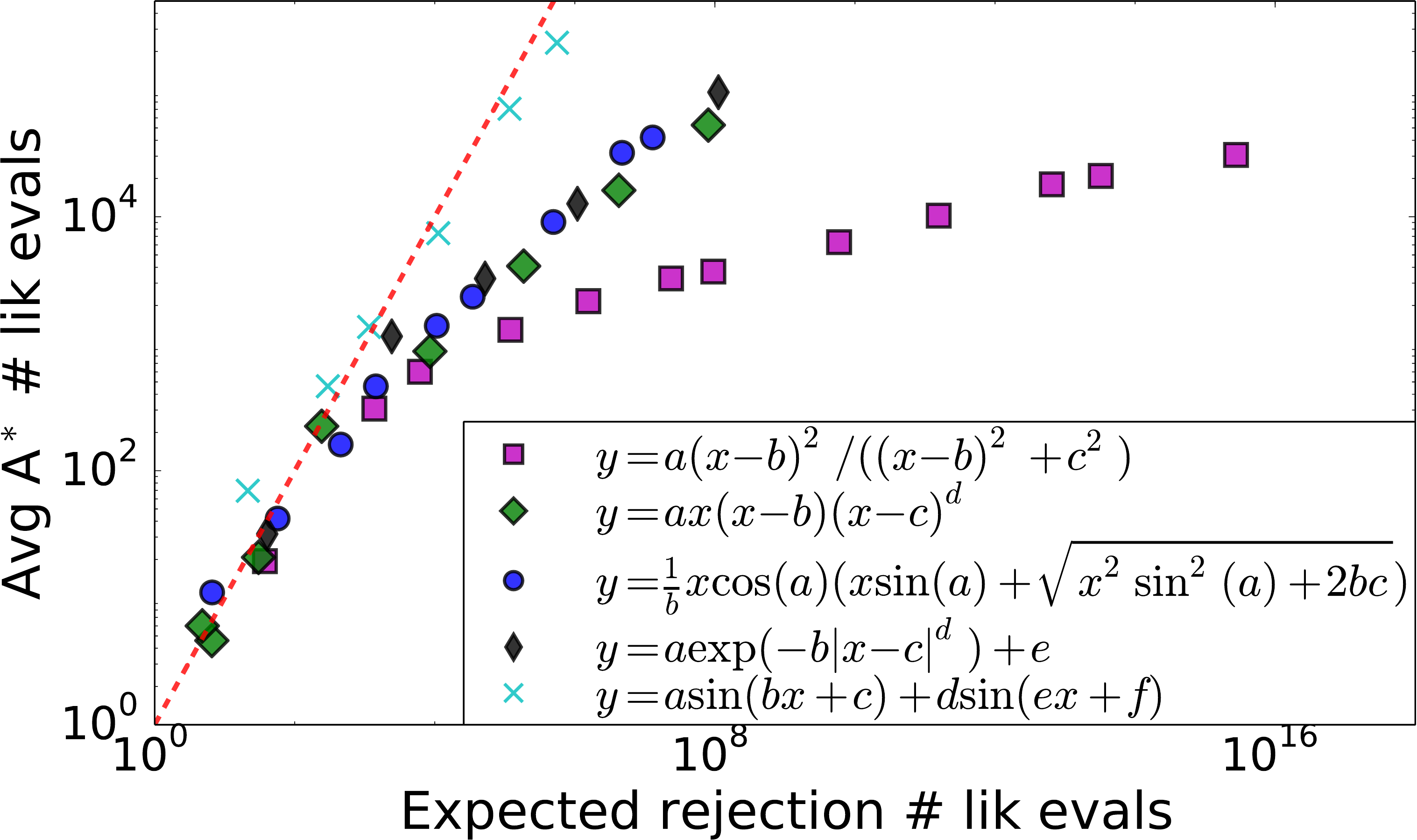} \\
\footnotesize{(a) vs.\ peakiness }&
\hspace{7pt}
\footnotesize{(b) vs.\ \# pts }&
\footnotesize{(c) Problem-dependent scaling}
  \\
\end{tabular}
\caption{
\footnotesize{
(a) Drill down algorithm performance on $p(x) = \exp(-x)/(1 + x)^a$
as function of $a$.
(b) Effect of different bounding strategies as a function of number
of data points; number of likelihood and bound evaluations are reported.
(c) Results of varying observation noise in several nonlinear regression
problems.
}
}
\label{fig:regression}
\end{figure}

\vspace{-7pt}
\subsection{Scaling versus Peakiness and Dimension}
\vspace{-7pt}

In the first experiment, we sample from 
$p(x) = \exp(-x)/(1 + x)^a$ for $x>0,a>0$ using
$\exp(-x)$ as the proposal distribution. 
In this case, $o(x) = -a \log(1 + x)$ which is
unimodal, so the drill down variant of \ouralgorithm~can be used.
As $a$ grows, the function
becomes peakier; while this presents significant difficulty
for vanilla rejection sampling, the cost to \astar~is just the cost of
locating the peak, which is essentially binary search. Results averaged
over 1000 runs appear
in \figref{fig:regression} (a). 

In the second experiment, we run \ouralgorithm~on the clutter problem
\cite{minka2001expectation}, which estimates the mean of a fixed covariance isotropic
Gaussian under the assumption that some points are outliers.  We put a
Gaussian prior on the inlier mean and set $i(x)$ to be equal to the prior,
so $o(x)$ contains just the likelihood terms.
To compute bounds on the total log likelihood, we compute upper bounds 
on the log likelihood of each point independently then sum up these bounds.
We will refer to these as ``constant'' bounds.
In $D$ dimensions, we generated 20 data points with half within $[-5,-3]^D$
and half within $[2,4]^D$, which ensures that the posterior is sharply bimodal, 
making vanilla MCMC quickly inappropriate as $D$ grows.
The cost of drawing an exact sample as a function of $D$ (averaged over 100 runs)
grows exponentially in $D$, but the problem
remains reasonably tractable as $D$ grows ($D=3$ requires 900 likelihood
evaluations, $D=4$ requires 4000). The analogous \osstar~algorithm run on
the same set of problems requires $16\%$ to $40\%$ more computation
on average over the runs.

\vspace{-7pt}
\subsection{Bounding Strategies}
\label{sec:bounding_strategies}
\vspace{-7pt}

Here we investigate alternative strategies
for bounding $o(x)$ in the case where $o(x)$ is a sum of
per-instance log likelihoods.
To allow easy implementation of a variety of bounding strategies, 
we choose the simple problem of estimating the mean of a 1D Gaussian
given $N$ observations.
We use three types of bounds: constant bounds as in the clutter problem;
linear bounds, where we compute linear upper bounds on each term of the sum,
then sum the linear functions and take the max over the region;
and quadratic bounds, which are the same as linear except quadratic bounds are computed
on each term. In this problem, quadratic bounds are tight.
We evaluate \ouralgorithm~using each of the bounding
strategies, varying $N$. See \figref{fig:regression} (b) for results.

For $N=1$, all bound types are equivalent when each expands around the same point.  For larger $N$, the looseness of each per-point bound becomes important.
The figure shows that, for large $N$, using linear bounds multiplies the number of evaluations by 3, compared to tight bounds.
Using constant bounds multiplies the number of evaluations by $O(\sqrt{N})$.
The Appendix explains why this happens and shows that this behavior is expected for any estimation problem where the width of the posterior shrinks with $N$.

\vspace{-7pt}
\subsection{Using Generic Interval Bounds}
\vspace{-7pt}

Here we study the use of bounds that are derived automatically by means
of interval methods \cite{hansen2003global}. This 
suggests how \ouralgorithm~(or \osstar)
could be used within a more general purpose
probabilistic programming setting. We chose a number of nonlinear regression
models inspired by problems in physics, computational ecology, and biology.
For each, we use  FuncDesigner \cite{funcdesigner} 
to symbolically construct $o(x)$ and automatically compute the bounds
needed by the samplers.

Several expressions for $y=f(x)$ appear in the legend of \figref{fig:regression} (c), where
letters $a$ through $f$ denote parameters that we wish
to sample. The model in all cases
is $y_n=f(x_n) + \epsilon_n$ where $n$ is the data point index and $\epsilon_n$ is Gaussian noise.
We set uniform priors from a reasonable range for all parameters (see Appendix) and generated
a small (N=3) set of training data from the model so that posteriors are multimodal.
The peakiness of the posterior can be controlled by the magnitude of the observation noise; we varied
this from large to small to produce problems over a range of difficulties.
We use \ouralgorithm~to sample from the posterior five times for each model and noise setting
and report the average number of likelihood evaluations needed in \figref{fig:regression} (c) (y-axis).
To establish the difficulty of the problems, we estimate the expected number of likelihood
evaluations needed by a rejection sampler to accept a sample.
The savings over rejection sampling is often exponentially large, but it 
varies per problem and is not necessarily tied to the dimension. In the example
where savings are minimal, there are many symmetries in the model, which leads to
uninformative bounds. We also compared to \osstar~on the same class of problems. Here
we generated 20 random instances with a fixed intermediate observation noise value
for each problem and drew 50 samples, resetting the bounds after each sample. The
average cost (heuristically set to \# likelihood evaluations plus 2 $\times$
\# bound evaluations) of \osstar~for the five models in \figref{fig:regression} (c)
respectively
was 21\%, 30\%, 11\%, 21\%, and 27\% greater than for \astar.

\vspace{-7pt}
\subsection{Robust Bayesian Regression}
\label{sec:cauchy_regression}
\vspace{-7pt}

Here our aim is to do Bayesian inference in a robust linear regression model
$y_n =  \bw^\trans \bx_n + \epsilon_n$
where noise $\epsilon_n$ is distributed as standard Cauchy and $\bw$ has
an isotropic Gaussian prior. Given a dataset $\data =\{\bx_n, y_n\}_{n=1}^N$
our goal is to draw samples from the posterior $\prob(\bw \given \data)$.
This is a challenging problem because the heavy-tailed noise model
can lead to multimodality in the posterior over $\bw$.
The log likelihood is
$\loglik(\bw) = \sum_n \log( 1 + ( \bw^\trans \bx_n - y_n)^2)$.
We generated $N$ data points with input dimension $D$
in such a way that the posterior is bimodal and symmetric by setting
$\bw^* = [2, ..., 2]^\trans$, generating
$X' \sim \hbox{randn}(N/2,D)$ and $y' \sim X' \bw^* + .1 \times \hbox{randn}(N/2)$,
then setting 
$X = \left[  X'; X' \right]$
and 
$y = \left[ y'; -y' \right]$. 
There are then equally-sized modes near $\bw^*$ and $-\bw^*$.
We decompose the posterior into a uniform $\i(\cdot)$ within the interval $[-10,10]^D$
and put all of the prior and likelihood terms into $\o(\cdot)$.
Bounds are computed per point; in some regions the per point bounds are linear,
and in others they are quadratic. Details appear in the Appendix.

We compare to \osstar, using
two refinement strategies that are discussed in \cite{dymetman2012algorithm}.
The first is directly analogous to \ouralgorithm~and is the method we have
used in the earlier \osstar~comparisons.
When a point is rejected, refine the piece that was proposed from
at the sampled point, and split the dimension with largest side length.
The second method splits the region with largest probability under the proposal.
\begin{wrapfigure}[24]{R}[0pt]{0pt}
\begin{minipage}[t]{.5\textwidth}
\vspace{-25pt}
\begin{figure}[H]
\hspace{-10pt}
\begin{tabular}{cc}
\includegraphics[width=.5\columnwidth]{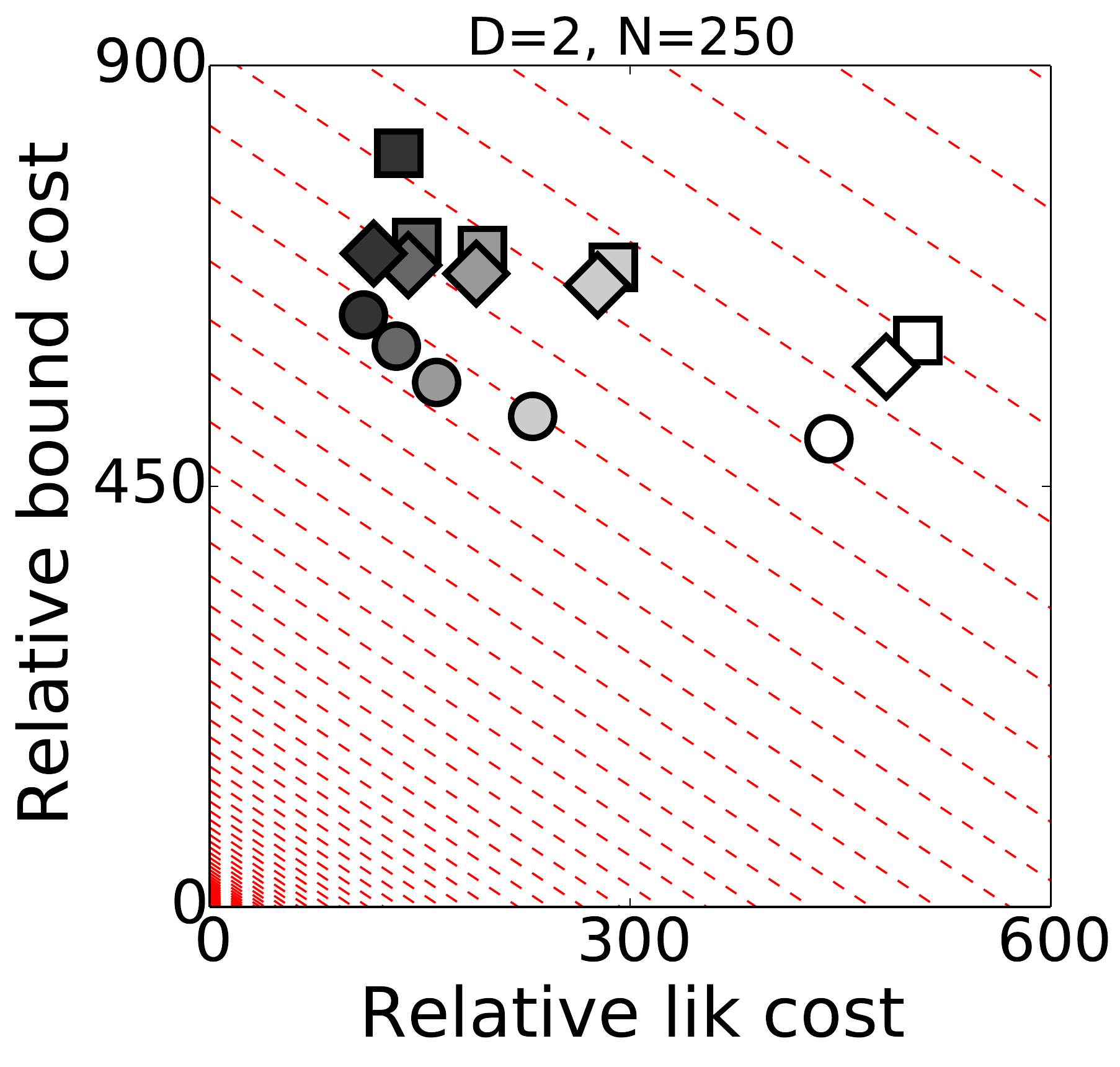} &
\hspace{-14pt}
\includegraphics[width=.52\columnwidth]{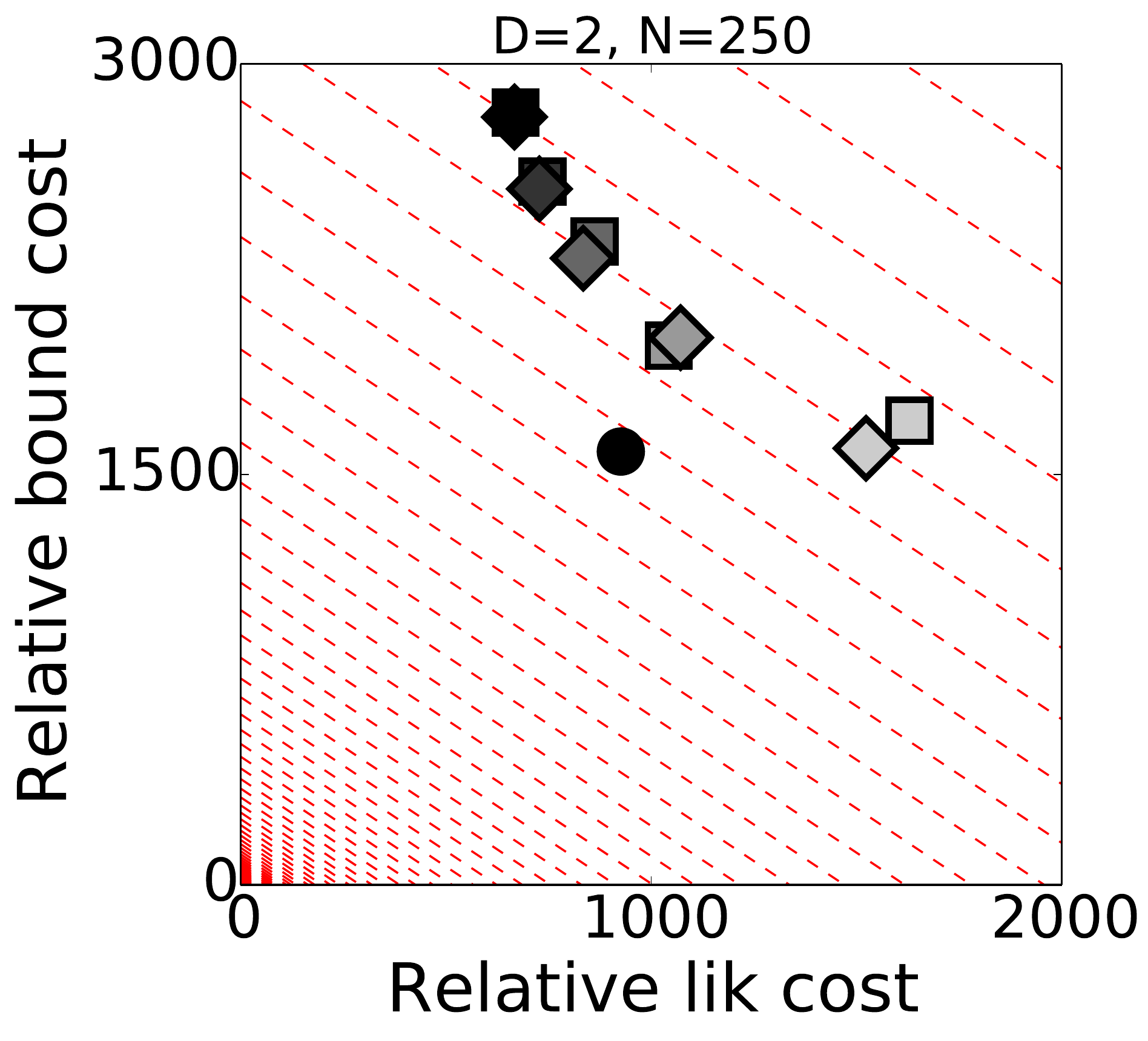}
\end{tabular}
\vspace{-10pt}
\caption{
\footnotesize{
\astar~(circles) versus \osstar~(squares and diamonds)
computational costs on Cauchy regression experiments
of varying dimension.
Square is refinement strategy that splits node where
rejected point was sampled; Diamond refines region with largest mass under
the proposal distribution. Red lines denote lines of equi-total computational
cost and are spaced on a log scale by 10\% increase increments.
Color of markers denotes the rate of
refinement, ranging from (darkest) refining for every rejection (for \osstar)
or one lower bound evaluation per node expansion (for \astar) to
(lightest) refining on 10\% of rejections (for \osstar) or performing 
$\hbox{Poisson}(\frac{1}{.1}-1) + 1$
 lower bound evaluations per node expansion (for \astar).
(left) Cost of drawing a single sample, averaged over 20 random data sets.
(right) Drawing 200 samples averaged over 5 random data sets.
Results are similar over a range of $N$'s and $D=1,\ldots,4$.
}
\label{fig:cauchy_results}
}
\end{figure}
\end{minipage}
\end{wrapfigure}
We ran experiments on several random draws of the data and report performance
along the two axes that are the dominant costs:\
how many bound computations were used, and how many
likelihood evaluations were used.  To weigh the tradeoff between the two, we 
did a rough asymptotic calculation of the costs of bounds versus likelihood computations
and set the cost of a bound computation to be $D+1$ times the cost of a likelihood computation.

In the first experiment, we ask each algorithm to draw a single exact sample
from the posterior. Here, we also report results for the variants of \ouralgorithm~and
\osstar~that trade off likelihood computations for bound computations as discussed
in \secref{sec:algs}. A representative result appears in \figref{fig:cauchy_results}
(left). Across operating points, \astar~consistently uses fewer bound evaluations and
fewer likelihood evaluations than both \osstar~refinement strategies.

In the second experiment, we ask each algorithm to draw 200 samples
from the posterior and experiment with the variants that reuse bound
information across samples. A representative result appears in \figref{fig:cauchy_results}
(right). Here we see that the extra refinement done
by \osstar~early on allows it to use fewer likelihood evaluations at
the expense of more bound computations, but \ouralgorithm~operates at
a point that is not achievable by \osstar.
For all of these problems, we ran a random direction slice sampler \cite{neal2003slice}
that was given 10 times the computational budget that \ouralgorithm~used to draw
200 samples.  The slice sampler had trouble mixing when $D>1$. Across the five runs for $D=2$, the
sampler switched modes once, and it did not ever switch modes when $D>2$.

\vspace{-7pt}
\section{Discussion}
\label{sec:discussion}
\vspace{-7pt}

This work answers a natural question: is there a Gumbel-Max trick for
continuous spaces, and can it be leveraged to develop tractable
algorithms for sampling from continuous distributions?  

In the discrete case, recent work on
``Perturb and MAP'' (\pam) methods \cite{Papandreou11, hazanNIPS2013,
tarlow2012randomized} that draw samples as the argmaxes of
random energy functions has shown value in developing approximate,
correlated perturbations. It is natural to think about 
continuous analogs in which exactness is abandoned in favor of 
more efficient computation. A question is if 
the approximations can be developed
in a principled way, like how \cite{hazan2012partition}
showed a particular form of correlated discrete perturbation gives rise to
bounds on the log partition function. Can analogous rigorous
approximations be established in the continuous case? 
We hope this work is a starting point for exploring that question.

We do not solve the problem of high dimensions. There are simple
examples where bounds become uninformative in high dimensions, such as
when sampling a density that is uniform over a hypersphere when using
hyperrectangular search regions. In this case, 
 little is gained over vanilla rejection
sampling.  An open question is if the split between $i(\cdot)$ and
$o(\cdot)$ can be adapted to be node-specific during the search. An
adaptive rejection sampler would be able to do this, which would allow
leveraging parameter-varying bounds in the proposal
distributions. This might be an important degree of freedom to
exercise, particularly when scaling up to higher dimensions.

There are several possible follow-ons including
the discrete version of \ouralgorithm~and evaluating
\ouralgorithm~as an estimator of the log partition function. 
In future work, we would like to explore taking advantage of conditional
independence structure to perform more intelligent search, hopefully
helping the method scale to larger dimensions. Example starting points
might be ideas from AND/OR search \cite{mateescu2007and} or branch and bound
algorithms that only branch on a subset of dimensions \cite{chandraker2008globally}.

\vspace{-7pt}
\section*{Acknowledgments}
\vspace{-7pt}
This research was supported by NSERC. We thank James Martens 
and Radford Neal for helpful discussions,
Elad Mezuman for help developing early ideas related to this work, 
and Roger Grosse for suggestions that greatly improved this work.

{\small
\bibliographystyle{unsrt}
\bibliography{biblio}
}

\newpage 
\section*{Appendix for ``\OurAlgorithm''}

In this appendix we prove the main theoretical results of the paper and provide
additional experimental details.
First, define the following shorthand
\begin{align*}
e_{\negenergy}(\g) &= \exp(-\g + \negenergy)\\
F_{\negenergy}(\g) &= \exp(-\exp(-\g + \negenergy))\\
f_{\negenergy}(\g) &= e_{\negenergy}(\g)F_{\negenergy}(\g)
\end{align*}
Thus $F_{\negenergy}(\g)$ is the CDF and $f_{\negenergy}(\g)$ the PDF of a $\GumbelDist(\negenergy)$. The following identities are easy to verify and will be reused throughout the appendix.
\begin{align}
 F_{\negenergy}(\g)F_{\gamma}(\g) &= F_{\log (\exp(\negenergy) + \exp(\gamma))}(\g)\\
\int_{x=a}^b e_{\gamma}(\g)F_{\negenergy}(\g) &= \left(F_{\negenergy}(b) - F_{\negenergy}(a)\right)\frac{\exp(\gamma)}{\exp(\negenergy)}
\end{align}

\section*{Joint Distribution of Gumbel Max and Argmax}

Suppose $\G(i) \sim \TruncGumbelDist(\negenergy(i), b)$ are $n$ independent truncated Gumbels and $Z = \sum_{i=1}^n \exp(\negenergy(i))$, then we are interested in deriving the joint distribution of $i^* = \argmax_{i=1}^n \G(i)$ and $\G(i^*) = \max_{i=1}^n \G(i)$.
\begin{align*}
p(k, g) &= p(i^*= k, \; G(i^*) = g) \\
&= p(\G(k) = g, \;  \G(k) \geq \max_{i \neq k} \G(i))\\
&= \frac{f_{\negenergy(k)}(g)\indicate{g \leq b}}{F_{\negenergy(k)}(b)} \prod_{i \neq k} \frac{F_{\negenergy(i)}(g)}{F_{\negenergy(i)}(b)} \\
&= \exp(-g + \negenergy(k)) \indicate{g \leq b} \prod_{i =1}^n \frac{F_{\negenergy(i)}(g)}{F_{\negenergy(i)}(b)}\\
&= \frac{\exp(\negenergy(k))}{Z} \exp(-g + \log Z) \indicate{g \leq b}  \prod_{i =1}^n \frac{F_{\negenergy(i)}(g)}{F_{\negenergy(i)}(b)}\\
&= \frac{\exp(\negenergy(k))}{Z}\frac{f_{\log Z}(g)\indicate{g \leq b} }{F_{\log Z}(b)}
\end{align*}
This is the Gibbs distribution and the density of a $\TruncGumbelDist(\log Z, b)$. Thus, for any $B
\subseteq \{1 \dotdotdot n\}$
\begin{align*}
\max_{i \in B} \G(i)  &\sim \TruncGumbelDist\left(\log \sum_{i \in B} \exp(\negenergy(i)),\;b\right)\\
\argmax_{i \in B} \G(i)  &\sim \frac{\indicate{i \in B}\exp(\negenergy(i))}{\sum_{i \in B} \exp(\negenergy(i))} \\
\max_{i \in B} \G(i)   \; &\perp \; \argmax_{i \in B} \G(i) 
\end{align*}
These results are well-known. The fact that max Gumbel value has a location that is the log partition
function means we can use samples of it as an estimator of log partition functions with known variance $\pi^2/6N$ for $N$ samples \cite{gumbel}. \eqref{eq:luceschoice} shows that Gumbels satisfy Luce's choice
axiom \cite{YellottJr1977109}. In fact, it is a well-known result in random choice theory that the only distribution satisfying \eqref{eq:luceschoice} is Gumbel. Notice that the argmax is also independent of the bound, and $b = \infty$ is a valid choice.

\clearpage

\section*{Analysis of Top-Down Construction}

\subsection*{Correctness of the Top-Down Construction of the \OurProcess}

\begin{figure}[t]
\begin{algorithm}[H]
\caption{\footnotesize{In-Order Construction}}
\label{alg:gupinorder}
\begin{algorithmic}
   \INPUT{sample space $\Omega$, sigma-finite measure $\mu(B)$}
   \STATE $\G_{1} \sim \GumbelDist(\log \measure(\Omega)) $
   \STATE $\X_1 \sim \exp(\negenergy(\x)) / \measure(\Omega)$
   \STATE $k \gets 1$
   \WHILE {$\Omega \neq \emptyset$}
	\STATE $k \gets k + 1$
	\STATE $\Omega \gets \Omega - \{\X_{k-1}\}$
   	\STATE $\G_{k} \sim \TruncGumbelDist(\log \measure(\Omega), \G_{k-1}) $
  	\STATE $\X_{k} \sim \indicate{\x \in \Omega} \exp(\negenergy(\x)) /  \measure(\Omega)$
	\STATE {\bf yeild} $\, (\G_k, \X_k)$
   \ENDWHILE
\end{algorithmic}
\end{algorithm}
\end{figure}

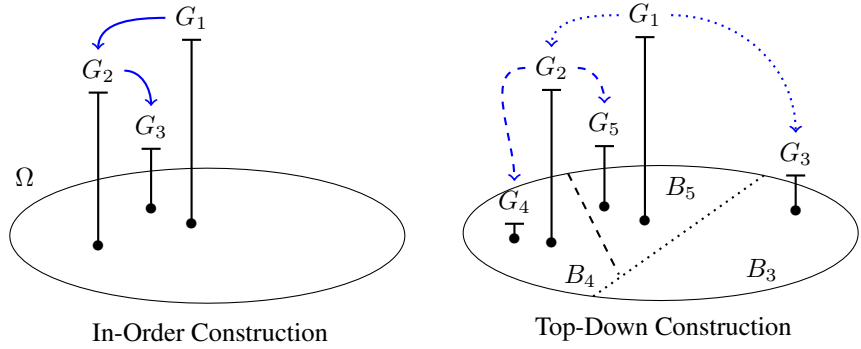
\begin{figure}[t]
\begin{center}
\tdplotsetmaincoords{70}{110}
\begin{tikzpicture}[scale=1.75,
			    tdplot_main_coords,
			    smooth
]
\draw (0,0,0) circle (1.5);
\node at (-0.2,-0.2,0) (g1) {$\bullet$};
\draw[thick,-|] (g1.center) -- ($(g1.center) + (0,0, 1.5)$) node[above](g1lab){$\G_1$};

\node at (0.5,-0.7,0) (g2) {$\bullet$};
\draw[thick,-|] (g2.center) -- ($(g2.center) + (0,0, 1.25)$) node[above](g2lab){$\G_2$};

\node at (-0.4,-0.6,0) (g3) {$\bullet$};
\draw[thick,-|] (g3.center) -- ($(g3.center) + (0,0, 0.5)$) node[above](g3lab){$\G_3$};

\draw [->,color=blue, thick] (g1lab.west) to [out=180,in=90] (g2lab.north);

\draw [->,color=blue,  thick] (g2lab.east) to [out=0,in=90] (g3lab.north);

\node at (-0.75,-1.75) {$\Omega$};
\node at (2, 0.75,0) {In-Order Construction};

\end{tikzpicture}
\begin{tikzpicture}[scale=1.75,
			    tdplot_main_coords,
			    smooth
]
\draw (0,0,0) circle (1.5);
\draw [dotted, thick] (1.5, 0,0) -- (-1.486, 0.3, 0);
\draw [dashed, thick] (-1,-1.112, 0) -- ($((1.4697, 0.3, 0) + 0.2*(-2.4697, -1.412)$);
\node at (-0.2,-0.2,0) (g1) {$\bullet$};
\draw[thick,-|] (g1.center) -- ($(g1.center) + (0,0, 1.5)$) node[above](g1lab){$\G_1$};

\node at (0.5,-0.7,0) (g2) {$\bullet$};
\draw[thick,-|] (g2.center) -- ($(g2.center) + (0,0, 1.25)$) node[above](g2lab){$\G_2$};

\node at (-0.8,0.8,0) (g3) {$\bullet$};
\draw[thick,-|] (g3.center) -- ($(g3.center) + (0,0, 0.3)$) node[above](g3lab){$\G_3$};

\node at (0.5,-1,0) (g4) {$\bullet$};
\draw[thick,-|] (g4.center) -- ($(g4.center) + (0,0, 0.13)$) node[above](g4lab){$\G_4$};

\node at (-0.4,-0.6,0) (g5) {$\bullet$};
\draw[thick,-|] (g5.center) -- ($(g5.center) + (0,0, 0.5)$) node[above](g5lab){$\G_5$};

\draw [->,color=blue, dotted, thick] (g1lab.west) to [out=180,in=90] (g2lab.north);
\draw [->,color=blue, dotted,  thick] (g1lab.east) to [out=0,in=90] (g3lab.north);

\draw [->,color=blue, dashed, thick] (g2lab.west) to [out=180,in=90] (g4lab.north);
\draw [->,color=blue, dashed,  thick] (g2lab.east) to [out=0,in=90] (g5lab.north);

\node at (2, 0.75,0) {Top-Down Construction};
\node at (0.5, 1,0) {$B_3$};
\node at (-1, -0.2,0) {$B_5$};
\node at (1.1, -0.25,0) {$B_4$};

\end{tikzpicture}
\end{center}
\caption{Visualization of a realization of a Gumbel process as produced by the Top-Down (Alg. \ref{alg:gup}) and In-Order  (Alg. \ref{alg:gupinorder}) constructions for the first few steps. Blue arrows indicate truncation. Black lines in $\Omega$ indicate partitioning for \algref{alg:gup}. In particular $B_2 = B_4 \cup B_5$. Note the sense in which they are simply re-orderings of each other.}
\label{fig:gup}
\end{figure}

The goal of this section is to prove that the Top-Down Construction constructs the Gumbel process. In particular, we will argue if we run \algref{alg:gup} with $\measure$ on $\Omega$, then the collection
\begin{align*}
\GUP^{\prime}_{\volume} = \{\max \{\G_k \given \X_k \in B\} \given B \subseteq \Omega\}
\end{align*}
is a Gumbel process $\GUP^{\prime}_{\volume} \overset{d}{=} \GUP_{\volume}$. In order to do this we consider a special case of \algref{alg:gup} in which space is not subdivided, $partition(B) = (B, \emptyset)$. In this case the construction takes on a particular simple form, since no queue is needed, see \algref{alg:gupinorder}. We call this special case the In-Order Construction, because is produces the Gumbel values in non-increasing order. 

We proceed by arguing that subdividing space has no effect on the distribution of the top $n$ Gumbels. This means that it would be impossible to distinguish a run of \algref{alg:gupinorder} from a run of \algref{alg:gup} with the Gumbel values sorted. This allows us to use any choice of $partition$ with a run of \algref{alg:gup} to analyze the distribution of $\max \{\G_k \given \X_k \in B\}$. More precisely
\begin{enumerate}
\item We argue that the top $n$ Gumbels of \algref{alg:gup} are distributed as in \algref{alg:gupinorder} regardless of $partition$. That is, if $[i]$ is the index of the $i$th largest Gumbel, then for $1 \leq n \leq |\Omega|$
\begin{align*}
&\G_{[1]} \sim \GumbelDist(\log \volume(\Omega))\\
&\G_{[k]} \sim \TruncGumbelDist(\log \volume(\Omega_{k-1}), \G_{[k-1]}) \text{ for } 1 < k \leq n\\
&\X_{[k]} \sim \indicate{\x \in \Omega_k}\exp(\negenergy(x))/\volume(\Omega_k) \text{ for } 1 \leq k \leq n
\end{align*}
This implies that the distribution over $\{\max \{\G_k \given \X_k \in B\} \given B \subseteq \Omega\}$ is invariant under the choice of $partition$ function in \algref{alg:gup}.
\item We derive the following for a specific choice of $partition$
\begin{align*}
\max\{ \G_k \given \X_k \in B\} &\sim \GumbelDist(\log \volume(B))\\
\max\{ \G_k \given \X_k \in B^c\} &\sim \GumbelDist(\log \volume(B^c))\\
\max\{ \G_k \given \X_k \in B\} &\;\perp\;\max\{ \G_k \given \X_k \in B^c\}
\end{align*}
By the previous result, this is the distribution for any choice of $partition$ (provided it doesn't produce immeasurable sets) giving us conditions 1. and 2. of Definition \ref{def:gup}. Condition 3. is easily satisfied.
\end{enumerate}
This proves the existence of the Gumbel process.

\subsubsection*{Equivalence Under $partition$}

We will proceed to show that the distribution over $\{\max \{\G_k \given \X_k \in B\} \given B \subseteq \Omega\}$ is invariant under the choice of $partition$ function. To do so we argue that the top $n$ Gumbels from \algref{alg:gup} all have the distribution from \algref{alg:gupinorder}. That is, if $[k]$ is the index of the $k$th smallest Gumbel in the tree from \algref{alg:gup} and $\Omega_k = \Omega - \cup_{i=1}^{k-1}\{\X_{[k]}\}$. Then for all $n \leq |\Omega|$
\begin{align*}
&\G_{[1]} \sim \GumbelDist(\log \volume(\Omega))\\
&\G_{[k]} \sim \TruncGumbelDist(\log \volume(\Omega_{k-1}), \G_{[k-1]}) \text{ for } 1 < k \leq n\\
&\X_{[k]} \sim \indicate{\x \in \Omega_k}\exp(\negenergy(x))/\volume(\Omega_k) \text{ for } 1 \leq k \leq n
\end{align*}
Notice that whenever $\measure(\Omega_k) = \measure(\Omega_{k+1})$ we can omit the removal of $\X_k$ and still have the same distribution. In the case of continuous $\measure$ we can completely omit all removals and set $\Omega_k = \Omega$.
\begin{proof}
We proceed by induction. For $n=1$, clearly
\begin{align*}
&\G_{[1]} = \G_1 \sim \GumbelDist(\log \volume(\Omega))\\
&\X_{[1]} = \X_1 \sim \exp(\negenergy(x))/\volume(\Omega)
\end{align*}
Now for $1 < n \leq |\Omega|$, consider the the top $n$ nodes from a single realization of the process. Let $[< \!\!k] = \{[1], \dotdotdot [k-1]\}$, the indices of the first $k-1$ Gumbels. By the induction hypothesis we know their distribution and they form a partial tree of the completely realized tree. 
Our goal is to show that
\begin{align*}
&\G_{[n+1]} \sim \TruncGumbelDist(\log \volume(\Omega_{n+1}), \G_{[n]})\\
&\X_{[n+1]} \sim \indicate{x \in \Omega_{n+1}}\exp(\negenergy(x))/\volume(\Omega_{n+1})
\end{align*}
The boundary of the max partial tree are the nodes $i$ that are on the Queue and have not been expanded. We know that conditioned on $[< \!\!n+1]$ that $\G_{[n+1]} = \max_{i \notin [< n+1]} \G_i$ will come from this boundary, i.e. $\G_{[n+1]} = \max_{i \in boundary} \G_i$. The first step is to realize that the sets $B_i$ on the boundary of the max partial tree form a partition of $\Omega_{n+1}$. If $\g_{[n+1]} = \max_{i \in boundary} \g_i$ and $p_i$ is the parent of node $i$, then
\begin{align*}
&p(\forall i \in boundary, \G_i = \g_{i}, \G_{[1]} = \g_{[1]} \dotdotdot \G_{[n]} = \g_{[n]} \given [< \!\!n+1])\\
&\propto \prod_{i \in boundary} f_{\log B_i} (g_i) \indicate{g_{p_i} > g_i} \prod_{k = 1}^n  f_{\log \measure(\Omega_k)}(\g_{[k]}) \indicate{\g_{[k]} > \g_{[k+1]}}
\end{align*}
Because products of indicator functions are like intersections
\begin{align*}
p(\forall i \in boundary, \G_i = \g_{i} \given \G_{[1]} = \g_{[1]} \dotdotdot \G_{[n]} = \g_{[n]}, [< \!\! n+1]) \propto \prod_{i \in boundary} f_{\log \measure(B_i)}(\g_i) \indicate{\G_{[n]} > \g_i}
\end{align*}
In other words, the boundary Gumbels are independent and $\G_i \sim \TruncGumbelDist(\log \measure (B_i), \G_{[n]})$. Notice that the subsets of the boundary form a complete partition of $\Omega_{n+1}$, thus we get
\begin{align*}
\G_{[n+1]} \sim \TruncGumbelDist(\log \volume(\Omega_{n+1}), \G_{[n]})
\end{align*}
The location $\X_{[n+1]}$ has the following distribution:
\begin{align*}
[n+1] \sim \argmax \{\G_i \given i \in boundary\}\\
\X_{[n+1]} \sim \indicate{\x \in B_{[n+1]}} \exp(\negenergy(\x))/\measure(B_{[n+1]})
\end{align*}
Again, because the $B_i$ is a partition of $\Omega_{n+1}$, this is a mixture distribution in which subsets $B_{[n+1]}$ are sampled with probability $\measure(B_{[n+1]})/ \measure(\Omega_{n+1})$ and then $\X_{[n+1]}$ is sampled from $\indicate{\x \in B_{[n+1]}}\exp(\negenergy(\x))/\measure(B_{[n+1]})$. Thus,
\begin{align*}
\X_{[n+1]} \sim \indicate{\x \in \Omega_{n+1}}\exp(\negenergy(x))/\volume(\Omega_{n+1})
\end{align*}
and
by the independence of the max and argmax we get that $\X_{[n+1]}$ is independent of $\G_{[n+1]}$.
\end{proof}

\subsubsection*{Joint Marginal of Max-Gumbels in $B$ and $B^c$} Because the joint distribution over the entire collection $\{\max\{ \G_k \given \X_k \in B\} \given B \subseteq \Omega\}$ is the same regardless of $partition$, this implies that the joint of $\max\{ \G_k \given \X_k \in B\}$ and $\max\{ \G_k \given \X_k \in B^c\}$ for any specific choice of partition is indeed the joint marginal for \emph{any} $partition$. In particular we show
\begin{align*}
\max\{ \G_k \given \X_k \in B\} &\sim \GumbelDist(\log \volume(B))\\
\max\{ \G_k \given \X_k \in B^c\} &\sim \GumbelDist(\log \volume(B^c))\\
\max\{ \G_k \given \X_k \in B\} &\;\perp\;\max\{ \G_k \given \X_k \in B^c\}
\end{align*}
\begin{proof}
Consider the $partition$ that first partitions $\Omega$ into $B$ and $B^c$. In this case we consider the distribution over $\G_{B} = \max\{ \G_k \given \X_k \in B\}$ and $\G_{B^c}  = \max\{ \G_k \given \X_k \in B^c\}$ in \algref{alg:gup}. If $\X_1 \in B$, then $\G_B = \G_1$ and $\G_{B^c} = \G_3$. Otherwise $\G_{B^c} = \G_1$ and $\G_{B} = \G_2$. Thus, $\G_B > \G_{B^c}$ iff $\X_1 \in B$. Using this knowledge we can split the distribution over $\G_B$ and $\G_{B^c}$ into two events.
\begin{align*}
&p(\G_B = g_b, \G_{B^c} = g_{B^c}) \\
&= p(\G_B = g_b, \G_{B^c} = g_{B^c} \given \G_B > \G_{B^c})p(\G_B > \G_{B^c}) + p(\G_B = g_b, \G_{B^c} = g_{B^c} \given \G_B \leq \G_{B^c})p(\G_B \leq \G_{B^c})\\
&=f_{\log \measure(\Omega)}(g_{B})\frac{f_{\log \measure(B^c)}(g_{B^c})\indicate{g_{B} > g_{B^c}}}{F_{\log \measure (B^c)}(g_B)}\frac{\measure(B)}{\measure(\Omega)} + f_{\log \measure(\Omega)}(g_{B^c})\frac{f_{\log \measure(B)}( g_{B})\indicate{g_{B} \leq g_{B^c}}}{F_{\log \measure (B)}(g_{B^c})}\frac{\measure(B^c)}{\measure(\Omega)} \\
&= f_{\log \measure(B)}(g_{B}) f_{\log \measure(B^c)}(g_{B^c})\indicate{g_{B} > g_{B^c}} + f_{\log \measure(B^c)}(g_{B^c}) f_{\log \measure(B)}(g_{B})\indicate{g_{B} \leq g_{B^c}}\\
&= f_{\log \measure(B)}(g_{B}) f_{\log \measure(B^c)}(g_{B^c})
\end{align*}
This is the density of two independent Gumbels with locations $\log \measure (B)$ and $\log \measure (B^c)$. This proves our result.
\end{proof}

\section*{Analysis of \OurAlgorithm}

This section deals with the correctness and termination of \ouralgorithm. We exclusively analyze the continuous version of \ouralgorithm. Recall we have two continuous measures $\measure(B) = \int_{x \in B} \exp(\negenergy(x))$ and $\measurei(B) = \int_{x \in B} \exp(\i(x))$ such that we can decompose $\phi(x)$ in a tractable $i(x)$ and intractable but boundable component $o(x)$.
\begin{align*}
\phi(x) = i(x) + o(x)
\end{align*}

\subsection*{Termination of \OurAlgorithm}

In this section we argue that \ouralgorithm~terminates with probability one by bounding it with the runtime of global-bound \ouralgorithm. We analyze global-bound \astar~more closely. 

Consider running \astar~with two different sets of bounds on the same realization of the Gumbel process.  The returned sample, the final lower bound, and the split chosen for any region will be the same.  The only thing that changes is the set of nodes in the tree that are explored. Let $U_{A^*}(B) = \G_k + M(B)$ be the upper bound at node $B$ for \astar~and $U(B) = \G_k + M$ be the upper bound at node $B$ for global-bound \astar. Because these algorithms are searching on the same realization we assume that $U(B) \geq U_{A^*}(B)$. Let $LB$ be the final lower bound---the optimal node. Because $U(B) \geq U_{A^*}(B)$, we know that  global-bound \astar~visits at least the nodes for which
\begin{align*}
U_{A^*}(B) \geq LB
\end{align*}
Finally, \astar~never visits nodes for which
\begin{align*}
U_{A^*}(B) < LB
\end{align*}
So, \astar~cannot visit a node that global-bound \astar~never visits. Thus, if global-bound \astar~terminates with probability one, then so does \astar. We now analyze the run time of global-bound \astar~more closely and discover a parallel with rejection sampling.

\begin{figure}[t]
\begin{algorithm}[H]
\caption{\footnotesize{Global-Bound \OurAlgorithm}}
\label{alg:topk}
    \begin{algorithmic}
\INPUT log density $i(x)$, difference $o(x)$, bounding function $M(B)$
\STATE $(LB, \; \X^* , \; k)\gets ( -\infty,\;  \text{null} ,\; 1)$
\STATE $\G_1 \sim (\GumbelDist(\log \measurei(\realsd))$
\STATE $\X_1 \sim \exp(i(\x)) / \measurei(\realsd))$
\STATE $\bound \gets M(\realsd)$
\WHILE {$LB < \G_k + \bound$}
		\STATE $LB_k \gets \G_k + o(\X_k)$
		\IF{$LB < LB_k$}
			\STATE $LB \gets LB_k$
			\STATE $X^* \gets \X_k$
		\ENDIF
	\STATE $k \gets k + 1$
   	\STATE $\G_{k} \sim \TruncGumbelDist(\log \measurei(\realsd), \G_{k-1}) $
  	\STATE $\X_{k} \sim \exp(\i(\x)) /  \measurei(\realsd)$
   \ENDWHILE
\OUTPUT $(LB, \X^{*})$
    \end{algorithmic}
\end{algorithm}
\end{figure}

\subsubsection*{Termination of Global-Bound \OurAlgorithm}

If a global bound $M \geq \o(\x)$ is reused at every node in \ouralgorithm, then it takes on a particularly simple form, \algref{alg:topk}; no queue is needed, and it simplifies to a search over the stream of $(\G_k, \X_k)$ values from the In-Order construction (Alg. \ref{alg:gupinorder}). Global-bound \ouralgorithm~is equivalent to rejection sampling. In particular, both rejection and \ouralgorithm~with constant bounds terminate after $k$ iterations with probability
\begin{align*}
(1-\rho)^{k-1} \rho
\end{align*}
where $\rho =  \measure(\realsd)(\exp(\bound) \measurei(\realsd))^{-1}$. 

Rejection terminates with this probability, because the termination condition is independent for each iteration. Thus, the distribution over the number of iterations is a geometric with probability:
\begin{align*}
\rho &= \prob\left(U \leq \exp(o(\X) - \bound)\right)\\
&= \expect[\exp(o(\X) - \bound)] \\
&= \frac{1}{\exp(\bound)}\int_{x \in \realsd} \exp(\negenergy(\x) - \i(\x)) \frac{\exp(\i(\x))}{\measurei(\realsd)}\\
&= \frac{\measure(\realsd)}{\exp(\bound) \measurei(\realsd)}
\end{align*}

That global-bound \astar~terminates with this probability is interesting, because the termination condition is not independent from the history of the Gumbel values. Nonetheless, the distribution over iterations is memoryless. Consider the stream of values $(\G_k, \X_k)$,
\begin{align*}
&\G_1 \sim \GumbelDist(\log \volume(\realsd))\\
&\G_k \sim \TruncGumbelDist(\log \volume(\realsd), \G_{k-1}) \text{ for } k > 1\\
&\X_k \sim \exp(\negenergy(\x))/\volume(\realsd)
\end{align*}
Global-bound \astar~terminates when $\max_{1 \leq i \leq k} \{\G_i + \o(\X_i)\} \geq \Gumbel_{k+1} + \bound$. In order to show that the distribution of $k$ is geometric we need simply to show that
\begin{align*}
\prob\left(\max_{1 \leq i \leq k}\{\G_i + \o(\X_i) - \bound\} < \Gumbel_{k + 1}\right) = (1 - \rho)^k
\end{align*}
\begin{proof}
First we show that the finite differences $D_k = G_{k+1} - G_{k}$ are mutual independent and 
\begin{align*}
-D_k \sim \Exponential(k)
\end{align*}
Inspecting the joint pdf for $\Gumbel_i$ with $1 \leq i \leq k + 1$
\begin{align*}
f(\gumbel_1, \ldots, \gumbel_{k+1}) &= f_{\mu}(\gumbel_1) \prod_{i=2}^{k+1} \frac{f_{\mu}(\gumbel_i) }{F_{\mu}(\gumbel_{i-1})}\indicate{g_i \leq g_{i-1} }\\
&= \left(\prod_{i=1}^{k+1} e_{\mu}(\gumbel_i)\right) F_{\mu}(\gumbel_{k+1}) \prod_{i=2}^{k+1}\indicate{g_i \leq g_{i-1} }
\end{align*}
We proceed with an inductive argument. First the base, with $k=1$, so for $d_1 < 0$
\begin{align*}
p(D_1 = d_1) &= \int_{g_1 = -\infty}^{\infty} e_{\mu}(\gumbel_1)  F_{\mu}(\gumbel_1 + d_1)\\
&= \exp(d_1)\int_{g_1 = -\infty}^{\infty}  e_{\mu}(\gumbel_1)  F_{\mu}(\gumbel_1)\\
&= \exp(d_1)\\
\end{align*}
Now by induction
\begin{align*}
\prob(D_1 \leq d_1, \ldots, D_k \leq d_k) &=  \int_{g_1 = -\infty}^{\infty} e_{\mu}(\gumbel_1)  \int_{g_2 = -\infty}^{\gumbel_1 + d_1}  e_{\mu}(\gumbel_2) \ldots \int_{g_{k+1}= -\infty}^{\gumbel_{k} + d_k}  f_{\mu}(\gumbel_{k+1})\\
 &= \int_{g_1 = -\infty}^{\infty} e_{\mu}(\gumbel_1)  \int_{g_2 = -\infty}^{\gumbel_1 + d_1}  e_{\mu}(\gumbel_2) \ldots \int_{g_{k}= -\infty}^{\gumbel_{k-1} + d_{k-1}} e_{\mu}(\gumbel_{k}) F_{\mu}(\gumbel_{k} + d_k)\\
 &= \exp(d_k) \prob(D_1 \leq d_1, \ldots, D_{k - 1} \leq d_{k-1} + d_k)\\
 &= \exp(d_k) \exp((k-1)(d_{k-1} + d_k)) \prod_{i=1}^{k-2} \exp(i d_i)\\
 &= \prod_{i=1}^{k} \exp(i d_i)\\
\end{align*}
Now we proceed to show that
\begin{align*}
\prob\left(\max_{1 \leq i \leq k}\{\G_i + \o(\X_i) - \bound\} < \Gumbel_{k + 1}\right) = (1 - \rho)^k
\end{align*}
First, let $Y = \o(\X) - M$ with $\X \sim \exp(\negenergy(\x))/\measure(\realsd)$ have PDF $h(y)$ and CDF $H(y)$. Notice that $\prob(Y \leq 0) = 1$ and $\expect[\exp(Y)] = \rho$. We rewrite the event of interest as a joint event in the finite differences of the Gumbel chain and then use the independence of the $X_i$
\begin{align*}
\prob\left(\max_{1 \leq i \leq k}\{\G_i + \o(\X_i) - \bound\} < \Gumbel_{k + 1}\right) &=\prob\left(\left\{\G_{k+1} - \G_i > \o(\X_i) - M\right\}_{i=1}^{k} \right)\\
&=\expect\left[\prob\left(\left\{\sum_{j = i}^k D_j > \o(\X_i) - M\right\}_{i=1}^{k} \;\middle|\;  \{ D_j\}_{j=1}^k \right)\right]\\
&=\expect\left[\prod_{i=1}^k H\left(\sum_{j=i}^k D_j\right) \right]
\end{align*}
looking more closely at this event
\begin{align*}
\mathbb{E} \left[\prod_{i=1}^k H\left(\sum_{j=i}^k D_j\right) \right] &= \int_{d_k = -\infty}^{0}\int_{d_{k-1} = -\infty}^{0} \ldots \int_{d_1 = -\infty}^0 \prod_{i=1}^k H\left(\sum_{j=i}^k d_j\right) \prod_{i=1}^k  i \exp(i d_i)\\
&= k! \int_{d_k = -\infty}^{0}\int_{d_{k-1} = -\infty}^{0} \ldots \int_{d_1 = -\infty}^0 \prod_{i=1}^k H\left(\sum_{j=i}^k d_j\right) \exp\left(\sum_{j=i}^k d_j\right)
\end{align*}
Now we do a sequence of tricky substitutions, $r_i = d_i + \sum_{j=i+1}^k d_j$ from $i = 1$ to $k$, and find that this integral equals
\begin{align*}
k! \int_{r_k = -\infty}^{0}\int_{r_{k-1} = -\infty}^{r_k} \ldots \int_{r_1 = -\infty}^{r_2} \prod_{i=1}^k H\left(r_i\right) \exp\left(r_i\right)
\end{align*}
Notice that this is basically an infinite triangle over a function that is symmetric in the $r_i$. Since it's multiplied by $k!$ it is equal to the sum over all the permutations of the $r_i$, which ends up giving us an infinite cube:
\begin{align*}
k! \int_{r_k = -\infty}^{0} \int_{r_{k-1} = -\infty}^{0} \ldots \int_{r_1 = -\infty}^{0} \prod_{i=1}^k H\left(r_i\right) \exp\left(r_i\right) = \left(\int_{r = -\infty}^{0} H\left(r\right) \exp\left(r\right) \right)^k
\end{align*}
All that remains is to evaluate $\int_{r = -\infty}^{0} H\left(r\right) \exp\left(r\right)$:
\begin{align*}
\int_{r = -\infty}^{0} H\left(r\right) \exp\left(r\right) &= \int_{r = -\infty}^{0}\int_{y = -\infty}^r h\left(y\right) \exp\left(r\right)\\
&= \int_{y = -\infty}^{0}\int_{r = y}^0 h\left(y\right) \exp\left(r\right)\\
&= \int_{y = -\infty}^{0} h(y) (1- \exp\left(y\right))\\
&= 1 - \rho
\end{align*}
and that completes the proof.
\end{proof}

\subsection*{Partial Correctness of \OurAlgorithm}
In this section we show that given termination the distribution returned by \astar~is correct. This depends only on the linearity result about the Gumbel process. We prove the linearity by using the auxiliary $\GUP_{\measurei}$ to measure the bounded difference $\o(\x)$.

At termination \astar~returns $LB$ and $\X^*$:
\begin{align*}
LB &= \max \{\G_k + \o(\X_k)\}\\
\X^* &= \argmax \{\G_k + \o(\X_k)\}
\end{align*}
where $(\G_k, \X_k)$ are a stream of Gumbels and locations obtained from running \algref{alg:gupinorder} with measure $\measurei$. Our goal is to show that $LB \sim \GumbelDist(\log \measure(\Omega))$ and $\X^* \sim \exp(\negenergy(\x))/\measure(\Omega)$. Thus, by Definition \ref{def:gup} it is sufficient to show that $\{\max \{\G_k + \o(\X_k) \given \X_k \in B\} \given B \subseteq \realsd\}$ is a \ourprocess~$\GUP_{\measure}$.

\begin{proof}
The correctness of the construction for $\GUP_{\measurei}$ implies the consistency and independence requirements. Thus we need only to verify the marginal of $\max \{\G_k + \o(\X_k) \given \X_k \in B\}$ is $\GumbelDist(\log \volume(B))$. To see why this is the case, consider a partition $p_1 \dotdotdot p_n$ of the range of $o(\x)$ and let 
\begin{align*}
P_j = \{ \x \; | \; p_{j-1} < o(\x) \leq p_{j}\}
\end{align*}
with $p_0 = -\infty$ and $p_n = \bound$. Then
\begin{align*}
\max \{\G_k + \o(\X_k) \given \X_k \in B\} = \max_j \{\max \{\G_k + \o(\X_k) \given \X_k \in B \cap P_j\} \}
\end{align*}
thus
\begin{align*}
\max_j \{\max \{\G_k \given \X_k \in B \cap P_j\}  + p_{j-1}\} \leq \max \{\G_k + \o(\X_k) \given \X_k \in B\} \leq \max_j \{\max \{\G_k \given \X_k \in B \cap P_j\}  + p_{j}\}
\end{align*}
$\{\G_k \given \X_k \in B \cap P_j\}  \sim \GumbelDist(\log \measurei(B \cap P_j))$, because $\G_k$ and $\X_k$ are samples from the process $\G_{\i}(B)$. Thus, \begin{align*}
\max_j\{\max \{\G_k \given \X_k \in B \cap P_j\}  + p_{j}\} \sim \GumbelDist(\log \sum_j \exp(\log \measurei (B \cap P_j) + p_j))
\end{align*}
similarly
\begin{align*}
\max_j \{\max \{\G_k \given \X_k \in B \cap P_j\}  + p_{j-1}\} \sim \GumbelDist(\log \sum_j \exp(\log \measurei (B \cap P_j) + p_{j-1}))
\end{align*}
we see that 
\begin{align*}
\sum_j \measurei (B \cap P_j)\exp(p_{j-1}) \rightarrow \int_{x \in B} \exp(i(x)) \exp(o(x)) \leftarrow  \sum_j \measurei (B \cap P_j)\exp(p_{j}) 
\end{align*}
as the partition gets finer. Since $ \int_{x \in B} \exp(i(x))\exp(o(x))= \measure(B)$ we get that
\begin{align*}
\max_j\{\max \{\G_k \given \X_k \in B \cap P_j\}  + p_{j}\} \overset{d}{\rightarrow} \GumbelDist\left(\log \measure(B)\right)\\
\max_j\{\max \{\G_k \given \X_k \in B \cap P_j\}  + p_{j-1}\} \overset{d}{\rightarrow} \GumbelDist\left(\log \measure(B)\right)
\end{align*}
Thus the distribution of $\max \{\G_k + \o(\X_k) \given \X_k \in B\}$ must be $\GumbelDist(\log \measure(B))$ and we're done.
\end{proof}

\subsection*{Explanation of Results from \secref{sec:bounding_strategies}}

Consider running \astar~with two
different sets of bounds on the same realization of the Gumbel
process.  The returned sample, the final lower bound, and the split
chosen for any region will be the same.  The only thing that changes
is the set of regions that are explored. 
Let $U_1(B)$ be an optimal
bound and $U_2(B)$ a suboptimal bound of region $R$.  The question is
how many more regions are explored by using $U_2$ instead of $U_1$.
Let $R^*$ be the region producing the returned sample.  Once this
region is explored, the lower bound reaches its final value and
\astar~will only explore regions with $U_2(B) > LB$.  These regions
will be new, have been unexplored by $U_1$, if $U_1(B) < LB$.  Before
$R^*$ is explored, \astar~will only explore regions with $U_2(B) >
U_2(R^*)$.  Since $U_2(R^*) > LB$, the condition that determines
whether a region $R$ is new is $U_2(B) > LB > U_1(B)$.

Suppose $R$ is explored using $U_1$, but its descendants are not.  How
many of its descendants are explored using $U_2$?  For the bounds we
consider in \secref{sec:bounding_strategies},
the suboptimality of $U_2$ is proportional to the region
width.  Let $B_d$ be the deepest descendant explored using $U_2$, and
let $n$ be the width of $R$ divided by $B_d$.  By the assumption on the
suboptimality of bounds being proportional to region width, we have
$(U_2(B)-U_1(B))/n > U_2(B_d) - U_1(B_d)$, and thus
$(U_2(B)-U_1(B))/n + U_1(B_d) > LB$.  This implies that $n$ is bounded
by a linear function of the suboptimality $U_2(B)-U_1(B)$.  The
balanced nature of the splitting process suggests that $B_d$ has depth
$\log_2(n)$ with high probability, so $n$ also bounds the total number
of explored descendants.  Therefore the total number of additional
regions explored by $U_2$ is linear in the total suboptimality of the
bounds of regions explored by $U_1$.

When the log-likelihood is a sum of $n$ terms and we apply a constant bound to
each term, the suboptimality of the total bound grows linearly with
$n$.  However, under the conditions of the Bernstein-von Mises
theorem, the posterior will concentrate around a peak of width
$O(n^{-1/2})$.  This shrinks the width of the significant regions,
reducing the suboptimality of an explored region to $O(\sqrt{n})$.
This is the trend we see in the plot.  If we apply a linear bound to
each term, the suboptimality is $O(({\rm width})^2)$ per term by the
Taylor remainder theorem, and overall $O(n ({\rm width})^2)$ which is
constant in $n$ for regions around the peak.  Therefore we expect
linear termwise bounds to explore a constant multiple of the number of
regions explored by optimal bounds.

\subsection*{Bounding the Cauchy Log Likelihood}

To perform inference in the Bayesian Robust Regression experiment, we need to upper bound 
$\max_{\bw \in [\lb{\bw}, \ub{\bw}]} \loglik(\bw)$ for each $\interval{\bw}$ encountered
in the search tree. For each region, we compute the bound in two
steps. First, for each $n$, compute the minimum and maximum possible values
of $d_n = \bw^\trans \bx_n - y_n$ using interval arithmetic \cite{hansen2003global},
yielding $\lb{d_n}$ and $\ub{d_n}$. For each $n$, we then construct a quadratic
bound on the Cauchy likelihood term $C(d) = -\log (1 + d^2)$ that is guaranteed to
be an upper bound so long as $d_n \in [\lb{d_n}, \ub{d_n}]$. The bound is
referred to as $B_n(d_n)$ and takes the form $B_n(d_n) = a_n d_n^2 + b_n d_n + c_n$.

The second derivative of the Cauchy likelihood
$C''(d)= \frac{4d^2}{(d^2+1)^2} - \frac{2}{d+1}$ changes sign only
twice, at $-1$ and $1$. Outside of $[-1,1]$, $C''(d)$ is positive (i.e., $C$ is convex), 
and inside it is negative (i.e., $C$ is concave). 
If an interval $\interval{d_n}$ is fully in a convex region, then we use a simple
linear bound of the line that passes through $(\lb{d_n}, C(\lb{d_n}))$ and $(\ub{d_n}, C(\ub{d_n}))$.
If an interval $\interval{d_n}$ is fully in a concave region and does not
contain $0$, then we use a linear bound that is tangent to $C$ at the midpoint 
of $\interval{d_n}$. If the interval contains any of $\{-1, 0, 1\}$ then we use
a quadratic bound. If the interval contains one of $-1$ or $1$, then we expand the
interval to include 0 and then proceed.
To compute the bound, we fix the bound function to have 
$B_n(0) = C(0)$ and $B'_n(0) = C'(0)$. Since the bound is quadratic, its second derivative
$2 a_n$ is constant. We set this constant to be the most negative value that ensures
that $B_n(d)$ is a valid bound over all of $\interval{d}$.
Concretely, the quadratic bound $B(d) = a d^2 + b d + c$ is constructed as follows. 
For each endpoint $d_{end} \in \{\lb{d_n}, \ub{d_n}\}$ that is not equal to 0, compute
\begin{align}
a & = \frac{C(d_{end}) - C(0)}{d_{end}^2}.
\end{align}
and choose the largest computed $a$.
Finally, solve for $b$ and $c$ to ensure that the derivative and value of the bound
match $C$ at 0:
\begin{align}
b & = f'(d_0) - 2a d_0 \\
c & = f(d_0) - a d_0^2 - b d_0.
\end{align}

\subsection*{Regression Experiment Priors}

All parameters were given uniform priors. The ranges are as follows:

$y = a \exp(-b\abs{x - c}^d) + e$

\begin{tabular}{ccc}
& a & [.1, 5] \\
& b & [.5, 5] \\
& c & [-5, 5] \\
& d & [.1, 5] \\
& e & [.1, 5] \\
\end{tabular}

$y = a \sin(b x + c) + d \sin(e x+f)$

\begin{tabular}{ccc}
& a & [-5, 5] \\
& b & [-5, 5] \\
& c & [-5, 5] \\
& d & [-5, 5] \\
& e & [-5, 5] \\
& f & [-5, 5] \\
\end{tabular}

$y = a (x - b)^2 / ((x-b)^2 + c^2)$

\begin{tabular}{ccc}
& a & [-5, 5] \\
& b & [-5, 5] \\
& c & [-5, 5] \\
\end{tabular}

$y = x  \frac{\cos(a) (x \sin(a) + \sqrt{x^2 \sin(a)^2 + 2bc})}{b}$
                                               
\begin{tabular}{ccc}
& a & [0.01, $\pi-.01$] \\
& b & [.1, 5] \\
& c & [0, 5] \\
\end{tabular}

$y = a x (x-b) (c-x)^d$

\begin{tabular}{ccc}
& a & [.01, 1] \\
& b & [.5, 1] \\
& c & [2, 3] \\
& d & [.1, 1] \\
\end{tabular}

\end{document}